%% file: EBLUPs-JASA-V2.tex
\documentclass[12pt]{article}
\usepackage{graphicx,psfrag,epsf}
\usepackage{enumerate}
\usepackage{natbib}
\usepackage{url} 

\RequirePackage{amsthm,amsmath,amsfonts,amssymb}
\RequirePackage[colorlinks,citecolor=blue,urlcolor=blue]{hyperref}
\usepackage{booktabs}
\usepackage{comment}

\usepackage{multicol}
\usepackage{multirow}
\usepackage{fancyhdr}
\usepackage{epsfig}
\usepackage{xparse}
\usepackage[english]{babel}															
\usepackage{caption}

\input{definition.tex}

\newcommand{\sumig}{\sum_{i=1}^g}
\newcommand{\sumjh}{\sum_{j=1}^h}
\newcommand{\sumkm}{\sum_{k=1}^m}
\newcommand{\yddd}{\bar{y}}
\newcommand{\yidd}{\bar{y}_{i.}}
\newcommand{\yijd}{\bar{y}_{ij}}
\newcommand{\ydjd}{\bar{y}_{.j}}
\newcommand{\xddd}{\bar{x}}
\newcommand{\xidd}{\bar{x}_{i.}}
\newcommand{\xijd}{\bar{x}_{ij}}
\newcommand{\xdjd}{\bar{x}_{.j}}
\newcommand{\eddd}{\bar{e}}
\newcommand{\eidd}{\bar{e}_{i.}}
\newcommand{\eijd}{\bar{e}_{ij}}
\newcommand{\edjd}{\bar{e}_{.j}}
\newcommand{\gammaid}{\bar{\gamma}_{i.}}
\newcommand{\gammadj}{\bar{\gamma}_{.j}}
\newcommand{\gammadd}{\bar{\gamma}}
\newcommand{\bxai}{\bx_{i}^{(a)}}
\newcommand{\bxait}{\bx_{i}^{(a)T}}
\newcommand{\bxbj}{\bx_{j}^{(b)}}
\newcommand{\bxbjt}{\bx_{j}^{(b)T}}
\newcommand{\bxabij}{\bx_{ij}^{(ab)}}
\newcommand{\bxabijt}{\bx_{ij}^{(ab)T}}
\newcommand{\bxwijk}{\bx_{ijk}^{(w)}}
\newcommand{\bxwijkt}{\bx_{ijk}^{(w)T}}

\newcommand{\barbxa}{\bar{\bx}^{(a)}}
\newcommand{\barbxb}{\bar{\bx}^{(b)}}
\newcommand{\barbxiab}{\bar{\bx}_{i.}^{(ab)}}
\newcommand{\barbxjab}{\bar{\bx}_{.j}^{(ab)}}
\newcommand{\barbxab}{\bar{\bx}^{(ab)}}
\newcommand{\barbxiw}{\bar{\bx}_{i.}^{(w)}}
\newcommand{\barbxjw}{\bar{\bx}_{.j}^{(w)}}
\newcommand{\barbxijw}{\bar{\bx}_{ij}^{(w)}}
\newcommand{\barbxw}{\bar{\bx}_{}^{(w)}}

\newcommand{\barbxat}{\bar{\bx}^{(a)T}}
\newcommand{\barbxbt}{\bar{\bx}^{(b)T}}
\newcommand{\barbxiabt}{\bar{\bx}_{i.}^{(ab)T}}
\newcommand{\barbxjabt}{\bar{\bx}_{.j}^{(ab)T}}
\newcommand{\barbxabt}{\bar{\bx}^{(ab)T}}
\newcommand{\barbxiwt}{\bar{\bx}_{i.}^{(w)T}}
\newcommand{\barbxjwt}{\bar{\bx}_{.j}^{(w)T}}
\newcommand{\barbxijwt}{\bar{\bx}_{ij}^{(w)T}}
\newcommand{\barbxwt}{\bar{\bx}_{}^{(w)T}}

\NewDocumentCommand{\barRia}{t^}{%
\IfBooleanTF{#1}
{\barRiaAux}
{\bar{R}_{i.}^{(a)}}%
}
\NewDocumentCommand{\barRiaAux}{m}{%
\bar{R}_{i.}^{(a)#1}%
}

\NewDocumentCommand{\barRjb}{t^}{%
\IfBooleanTF{#1}
{\barRjbAux}
{\bar{R}_{.j}^{(b)}}%
}
\NewDocumentCommand{\barRjbAux}{m}{%
\bar{R}_{.j}^{(b)#1}%
}

\NewDocumentCommand{\barRijab}{t^}{%
\IfBooleanTF{#1}
{\barRijabAux}
{\bar{R}_{ij}^{(ab)}}%
}
\NewDocumentCommand{\barRijabAux}{m}{%
\bar{R}_{ij}^{(ab)#1}%
}

\NewDocumentCommand{\barRijkw}{t^}{%
\IfBooleanTF{#1}
{\barRijkwAux}
{{R}_{ijk}^{(w)}}%
}
\NewDocumentCommand{\barRijkwAux}{m}{%
{R}_{ijk}^{(w)#1}%
}

\newcommand{\hsiga}{\hat{\sigma}_{\alpha}}  
\newcommand{\hsigb}{\hat{\sigma}_{\beta}}  
\newcommand{\hsiggama}{\hat{\sigma}_{\gamma}}  
\newcommand{\hsige}{\hat{\sigma}_{e}}

\newcommand{\blind}{1}

\begin{document}

\def\spacingset#1{\renewcommand{\baselinestretch}%
{#1}\small\normalsize} \spacingset{1}


\if1\blind
{
  \title{\bf Asymptotics for EBLUPs within crossed mixed effect models}
  \author{Ziyang Lyu\thanks{Ziyang Lyu, Email: \textsf{Ziyang.Lyu@unsw.edu.au}. S.A.~Sisson, Email: \textsf{Scott.Sisson@unsw.edu.au}. A.H.~Welsh \textsf {Alan.Welsh@anu.edu.au}. SAS and AHW are respectively supported by the Australian Research Council Discovery Projects DP220103269 and DP230101908. 
  }\hspace{.5cm} S.A.~Sisson\\
   UNSW Data Science Hub, and School of Mathematics and Statistics,\\
University of New South Wales\\
    A.H.~Welsh \\
   Research School of Finance, Actuarial Studies and Statistics,\\
Australian National University}
  \maketitle
} \fi

\if0\blind
{
  \bigskip
  \bigskip
  \bigskip
  \begin{center}
    {\LARGE\bf Asymptotics for EBLUPs within crossed mixed effect models}
\end{center}
  \medskip
} \fi
  \vspace{-0.9cm}
\begin{abstract}
In this article, we derive the joint asymptotic distribution of empirical best linear unbiased predictors (EBLUPs) for individual and cell-level random effects in a crossed mixed effect model. 
Under mild conditions (which include moment conditions instead of normality for the random effects and model errors), we demonstrate that as the sizes of rows, columns, and, when we include interactions, cells simultaneously increase to infinity, the distribution of the differences between the EBLUPs and the random effects satisfy central limit theorems.  These central limit theorems mean the EBLUPs asymptotically follow the convolution of the true random effect distribution and a normal distribution. Moreover, our results enable simple asymptotic approximations and estimators for the mean squared error (MSE) of the EBLUPs, which in turn facilitates the construction of asymptotic prediction intervals for the unobserved random effects. We show in simulations that our simple estimator of the MSE of the EBLUPs works very well in finite samples.  Finally, we illustrate the use of the asymptotic prediction intervals with an analysis of movie rating data.
\end{abstract}
\noindent%
\vfill

\newpage
\spacingset{1.45} 
\section{Introduction}
\label{sec:intro}
Mixed effect models are a comprehensive class of statistical models for the associations between response variables, observed covariates or explanatory variables, and unobserved random terms, known as random effects. Random effects are included in the model to accomplish various objectives: to represent sampling from a broader population, to incorporate dependence, to allow more concise models, and/or to enhance predictive accuracy. In some scenarios, such as the ranking problem presented in Section~\ref{sec7: real data}, the actual values of these random effects are important. In other scenarios (such as maximum likelihood or restricted maximum likelihood (REML) estimation) their distribution is important because a key step is integrating the unobserved random effects over their distribution.  In either case, it is good practice to estimate or predict the random effects and to analyze their distributions to obtain insights into the model's behavior and effectiveness.

In linear mixed models, the most widely used estimators of the realised values of the random effects are empirical best linear unbiased predictors (EBLUPs) which are estimated versions of best linear unbiased predictors (BLUPs) derived by \cite{henderson1959estimation}; see for example \cite{henderson1975best}, \cite{robinson1991blup} and  \cite{jiang1998asymptotic}.  EBLUPs have proved important in the prediction of breeding values \citep{henderson1984breeding}, small area means \citep{battese1988error, lyu2023small}, and in many other application areas.  Many predictions are presented as point predictions; approximations to their prediction mean squared error under normality developed by \cite{kackar1984approximations} and \cite{prasad1990estimation} are sometimes used as measures of uncertainty. \cite{chatterjee2008} provide an interesting discussion of the problem of setting prediction intervals.  Diagnostic methods using EBLUPs are discussed in, for example, \citet{lange1989diag}, \citet[ch 7]{verbeke2000lmm}, \citet[Sec 4.3.2]{pinheiro2000lmm}, \cite{nobre2007diag}, and 
\cite{schutzenmeister2012diag}.

Despite the extensive use of EBLUPs in practice, there are only limited results on their theoretical properties. \cite{jiang1998asymptotic} made significant contributions by demonstrating the consistency of EBLUPs for random effects in models with nested random effect structures. However, this analysis does not consider the asymptotic distribution of EBLUPs and it does not extend to models with crossed random effects.
\cite{lyu2019eblup,lyu2019estimation} explored the asymptotic distribution of EBLUPs in nested error regression models, but again did not address the complexities of crossed effects. This gap is noteworthy because crossed random effects are common in practical applications, including reader-based diagnostics \citep{withanage2015joint}, educational assessments \citep{menictas2023streamlined}, and customer ratings \citep{ghosh2022backfitting}.
The purpose of this paper is to address this gap by establishing the asymptotic properties of EBLUPs for models with crossed random effects.
 
\newcommand{\rddd}{\bar{r}}
\newcommand{\ridd}{\bar{r}_{i.}}
\newcommand{\rdjd}{\bar{r}_{.j}}
\newcommand{\rijd}{\bar{r}_{ij}}

We consider data arranged in a two-way table in which the cells correspond to the crossing of two categorical factors, hereinafter referred to as factor A (rows) and factor B (columns), respectively.   
Let \( [y_{ij}, \bx_{ij}^T]^T \) represent the  observed vector in cell \( (i,j) \), where \( y_{ij} \) is a scalar response variable and \( \bx_{ij} \) is a $p_0$-dimensional vector of covariates or explanatory variables, and $T$ denotes the transpose. 
The two-way crossed random effect model is
\begin{equation}\label{two-way cross model no interaction}
	y_{ij}=\mu(\bx_{ij})+\alpha_i+\beta_j+e_{ij}, \qquad i=1,\ldots,g, j=1,\ldots,h,
\end{equation}
where $\mu(\bx_{ij})$ is the conditional mean of the response given the covariate $\bx_{ij}$ (also called the regression function),
$\alpha_i$ is the random effect due to the $i$th row, $\beta_j$ is the random effect  due to the $j$th column, and $e_{ij}$ is the error term.  The unobserved variables $\alpha_i$, $\beta_j$ and $e_{ijk}$ are assumed to be independent and identically distributed with zero means and variances $\sigma_\alpha^2$, $\sigma_\beta^2$, and $\sigma_e^2$, respectively, and to be mutually independent.  We do not assume normality.  

In some studies, it is possible to replicate the design so that there are $m>1$ observations in each cell and, in such cases, we are able to include an additional interaction random effect in the model.  To accommodate this, we introduce an additional subscript $k$ to the data and consider the  two-way crossed random effect with interaction model 
\begin{equation}\label{two-way cross model}
	y_{ijk}=\mu(\bx_{ijk})+\alpha_i+\beta_j+\gamma_{ij}+e_{ijk},\quad i=1,\ldots,g,\, j=1,\ldots,h,\, k=1,\ldots,m,
\end{equation}
where $\gamma_{ij}$ is the interaction random effect  of row $i$ and column $j$, assumed to be independent and identically distributed with zero mean and variance $\sigma_\gamma^2$, independent of the other random terms. Of course, we can also fit the no interaction model ($\siggama^2=0$ and hence $\gamma_{ij}=0$) to these data; the results follow as a special case of considering the general model (\ref{two-way cross model}).  A further generalization to unequal numbers $m_{ij}$ of observations in each cell will not be considered here.

For the two-way crossed random effect with interaction model (\ref{two-way cross model}),
let $n = ghm$,  $\by=[y_{111},\ldots, y_{ghm}]^T$, $\bX=[\bx_{111},\ldots,\bx_{ghm}]^T$, $\balpha=[\alpha_1,\ldots,\alpha_g]^T$, $\bbeta=[\beta_1,$\ldots$,\beta_h]^T$, $\bgamma=[\gamma_{11},\ldots,\gamma_{gh}]^T$, and $\be=[e_{111},\ldots,e_{ghm}]^T$. 
The indices cycle through their possible values starting with the rightmost index first.
Then the matrix form of (\ref{two-way cross model}) is 
\begin{equation}\label{two-way matrix}
	\by=\mu(\bX)+\bZ\bu+\be,
\end{equation}
where $\bu=[\balpha^T,\bbeta^T,\bgamma^T]^T$ and $\bZ=[\bZ_1,\bZ_2,\bZ_3]$ with $\bZ_1=\bI_g\otimes\bone_h\otimes \bone_m$, $\bZ_2=\bone_g\otimes\bI_h\otimes \bone_m$ and	$\bZ_3=\bI_g\otimes\bI_h\otimes \bone_m$. Here we let $\bone_a$ denote the $a$-vector of ones, $\bI_a$ the $a \times a$ identity matrix, and $\otimes$ the Kronecker product. 
We let $\btheta=[\siga^2,\sigb^2,\siggama^2,\sige^2]^T$, $\bG(\btheta)=\var(\bu)=\diag(\siga^2\bI_g,\sigb^2\bI_h,\siggama^2\bI_{gh})$ and $\var(\be)=\sige^2\bI_n$. The variance-covariance matrix of $\by$, denoted by 
$\bV(\btheta)=\sige^2\bI_n+\bZ\bG(\btheta)\bZ^T=\bZ_1\bZ_1^T\sigma_\alpha^2+\bZ_2\bZ_2^T\sigma_\beta^2+\bZ_3\bZ_3^T\sigma_\gamma^2+\sige^2\bI_n,$
where
$\bZ_1\bZ_1^T=\bI_g\otimes\bJ_h\otimes \bJ_m$,  $\bZ_2\bZ_2^T=\bJ_g\otimes\bI_h\otimes \bJ_m$ and 
$\bZ_3\bZ_3^T=\bI_g\otimes\bI_h\otimes \bJ_m$, with $\bJ_a = \bone_a\bone_a^T$ an $a \times a$ matrix of ones.  Model (\ref{two-way cross model no interaction}) can be written in a similar form with $m=1$, and $\sigma_\gamma^2$, $\bgamma$ and $\bZ_3$ omitted.

The BLUPs for $\bu$ can be expressed as  $\bG^{-1}(\btheta)\bZ^T\bV^{-1}(\btheta)\{\by-\mu(\bX)\}$. To obtain exact expressions for the BLUPs, we can use the explicit expression for $\bV^{-1}(\btheta)$ obtained by following the algorithm of \cite{searle1979dispersion},   
	\begin{align*}
		\bV^{-1}(\btheta)=&
		\frac{1}{\lambda_0}(\bI_g\otimes\bI_h\otimes\bC_m)+\frac{1}{\lambda_1}(\bC_g\otimes\bC_h\otimes\bar{\bJ}_m)+\frac{1}{\lambda_2}(\bC_g\otimes\bar{\bJ}_h\otimes\bar{\bJ}_m)\\&
		+\frac{1}{\lambda_3}(\bar{\bJ}_g\otimes\bC_h\otimes\bar{\bJ}_m)+\frac{1}{\lambda_4}(\bar{\bJ}_g\otimes\bar{\bJ}_h\otimes\bar{\bJ}_m),
	\end{align*}
	where 
$\lambda_0=\sigma_e^2$, $\lambda_1=\sigma_e^2+m\sigma_\gamma^2$,
$\lambda_2=\sigma_e^2+m\sigma_\gamma^2+hm\sigma_\alpha^2$,
$\lambda_3=\sigma_e^2+m\sigma_\gamma^2+gm\sigma_\beta^2$, 
$\lambda_4=\sigma_e^2+m\sigma_\gamma^2+hm\sigma_\alpha^2+gm\sigma_\beta^2$, $\bar{\bJ}_a=\frac{1}{a}{\bJ}_a$ and $\bC_a=\bI_a-\bar\bJ_a$. 
This expression leverages the properties of the Kronecker product and is available for the balanced case of the crossed random effect with interaction model (\ref{two-way cross model})  and the crossed random effect with no interaction model (\ref{two-way cross model no interaction}). As far as we know, there is no explicit form for $\bV^{-1}$ in the general unbalanced case.
The BLUPs for $\alpha_i$, $\beta_j$, and $\gamma_{ij}$ in the model (\ref{two-way cross model}) are
\begin{align}
\hat\alpha_i(\mu,\btheta)&=\frac{hm\siga^2}{\lambda_2}\{\yidd-\frac{1}{hm}\sumjh\sumkm\mu(\bx_{ijk})\}-hm\siga^2(\frac{1}{\lambda_2}-\frac{1}{\lambda_4})\{\yddd-\frac{1}{n}\sumig\sumjh\sumkm\mu(\bx_{ijk})\},\nonumber
\\
\hat\beta_j(\mu,\btheta)&=\frac{gm\sigb^2}{\lambda_3}\{\ydjd-\frac{1}{gm}\sumig\sumkm\mu(\bx_{ijk})\}-gm\sigb^2(\frac{1}{\lambda_3}-\frac{1}{\lambda_4})\{\yddd-\frac{1}{n}\sumig\sumjh\sumkm\mu(\bx_{ijk})\},\nonumber
\\
\hat\gamma_{ij}(\mu,\btheta)&=\frac{m\siggama^2}{\lambda_1}\{\yijd-\frac{1}{m}\sumkm\mu(\bx_{ijk})\}-m\siggama^2(\frac{1}{\lambda_1}-\frac{1}{\lambda_2})\{\yidd-\frac{1}{hm}\sumjh\sumkm\mu(\bx_{ijk})\}\nonumber
\\&\quad-m\siggama^2(\frac{1}{\lambda_1}-\frac{1}{\lambda_3})\{\ydjd-\frac{1}{gm}\sumig\sumkm\mu(\bx_{ijk})\}\nonumber\\&\quad
+m\siggama^2(\frac{1}{\lambda_1}-\frac{1}{\lambda_2}-\frac{1}{\lambda_3}+\frac{1}{\lambda_4})\{\yddd-\frac{1}{n}\sumig\sumjh\sumkm\mu(\bx_{ijk})\},\label{BLUPs1}
\end{align}
with $\yddd=n^{-1}\sumig\sumjh\sumkm y_{ijk}$, $\yijd=m^{-1}\sumkm y_{ijk}$, $\yidd=(hm)^{-1}\sumjh\sumkm y_{ijk}$ and $\ydjd=(gm)^{-1}\sumig\sumkm y_{ijk}$.
We adopt the convention of using a dot as a placeholder subscript when required (as in $\yidd$ and $\ydjd$) but omit it when there is no ambiguity (as in $\yijd$ and $\yddd$).
The expressions in (\ref{BLUPs1}) show that for the BLUPs to converge in probability to their respective random effects $\alpha_i$, $\beta_j$, and $\gamma_{ij}$, a necessary condition is that $g, h, m\to\infty$.  The BLUPs under model (\ref{two-way cross model no interaction}) are given in (\ref{blups2}) below.

\newcommand{\bxais}{\bx_{i}^{*(a)}}
\newcommand{\bxaist}{\bx_{i}^{*(a)T}}
\newcommand{\bxbjs}{\bx_{j}^{*(b)}}
\newcommand{\bxbjst}{\bx_{j}^{*(b)T}}
\newcommand{\bxwijs}{\bx_{ij}^{*(w)}}
\newcommand{\bxwijst}{\bx_{ij}^{*(w)T}}

\newcommand{\barbxas}{\bar{\bx}^{*(a)}}
\newcommand{\barbxbs}{\bar{\bx}^{*(b)}}
\newcommand{\barbxiws}{\bar{\bx}_{i.}^{*(w)}}
\newcommand{\barbxjws}{\bar{\bx}_{.j}^{*(w)}}
\newcommand{\barbxws}{\bar{\bx}^{*(w)}}

\newcommand{\barbxast}{\bar{\bx}^{*(a)T}}
\newcommand{\barbxbst}{\bar{\bx}^{*(b)T}}
\newcommand{\barbxiwst}{\bar{\bx}_{i.}^{*(w)T}}
\newcommand{\barbxjwst}{\bar{\bx}_{.j}^{*(w)T}}
\newcommand{\barbxwst}{\bar{\bx}^{*(w)T}}

However, the mean function and the variance components in both models (\ref{two-way cross model no interaction}) and (\ref{two-way cross model})  are typically unknown and so must be estimated from data; the EBLUPs are the BLUPs evaluated at the maximum likelihood or REML estimators of the unknown parameters. The asymptotic properties of the EBLUPs depend on those of the estimators of the unknown parameters which, in crossed random effect models with linear regression functions were unknown \cite[Section 3.7.2]{jiang2017asymptotic} until 
\cite{lyu2024increasing} derived asymptotic results (as $g, h \rightarrow \infty$ for model (\ref{two-way cross model no interaction}) and $g, h, m \rightarrow \infty$ for model (\ref{two-way cross model})) for maximum likelihood and restricted or residual maximum likelihood (REML) estimators of these parameters.
We use these results to derive the asymptotic distribution of the EBLUPs and the cell-level EBLUPs for both models (\ref{two-way cross model no interaction}) and (\ref{two-way cross model}).   These results are more interesting  and  more difficult to derive than those for EBLUPs in nested models, because there are more random effects in the model, the EBLUPs have a more complicated form, and there are subtleties in the behavior of the sequences $g/h$, $g/m$ and $h/m$ to consider. 

The paper is structured as follows. Section~\ref{notation} introduces the notation. Sections~\ref{Sec3: Non Interaction Model} and~\ref{Sec4: Interaction Model} present central limit theorems for individual and cell-level EBLUPs under models (\ref{two-way cross model no interaction}) and (\ref{two-way cross model}) without assuming normality. These results allow us to write the asymptotic distributions of EBLUPs as the convolution of the true random effect distribution and a normal distribution. In Section~\ref{Sec4:MSE}, we provide approximations for the mean squared error (MSE) of the EBLUPs  and compare them with those of \citet{kackar1984approximations} and \citet{prasad1990estimation}. We discuss MSE estimation and construct asymptotic prediction intervals for random effects. Sections~\ref{Sec6:simulation} and ~\ref{sec7: real data} present a simulation study evaluating the finite sample performance of our MSE estimator, and
a real-data analysis of a movie rating dataset. Concluding remarks are given in Section~\ref{sec8: discussion}.

\newcommand{\cxias}{\bx_{i(c)}^{*(a)}}
\newcommand{\cxiast}{\bx_{i(c)}^{*(a)T}}
\newcommand{\cxiws}{\bar{\bx}_{i.(c)}^{*(w)}}
\newcommand{\cxiwst}{\bar{\bx}_{i.(c)}^{*(w)T}}
\newcommand{\cyias}{\bar{y}_{i.(c)}^*}

\newcommand{\cxjbs}{\bx_{j(c)}^{*(b)}}
\newcommand{\cxjbst}{\bx_{j(c)}^{*(b)T}}
\newcommand{\cxjws}{\bar{\bx}_{.j(c)}^{*(w)}}
\newcommand{\cxjwst}{\bar{\bx}_{.j(c)}^{*(w)T}}
\newcommand{\cyjbs}{\bar{y}_{.j(c)}^*}

\newcommand{\cxijws}{{\bx}_{ij(c)}^{*(w)}}
\newcommand{\cxijwst}{{\bx}_{ij(c)}^{*(w)T}}
\newcommand{\cyijws}{{y}_{ij(c)}^*}

\newcommand{\cxia}{\bx_{i(c)}^{(a)}}
\newcommand{\cxiat}{\bx_{i(c)}^{(a)T}}
\newcommand{\cxiab}{\bar{\bx}_{i.(c)}^{(ab)}}
\newcommand{\cxiabt}{\bar{\bx}_{i.(c)}^{(ab)T}}
\newcommand{\cxiw}{\bar{\bx}_{i.(c)}^{(w)}}
\newcommand{\cxiwt}{\bar{\bx}_{i.(c)}^{(w)T}}
\newcommand{\cyia}{\bar{y}_{i.(c)}}

\newcommand{\cxjb}{\bx_{j(c)}^{(b)}}
\newcommand{\cxjbt}{\bx_{j(c)}^{(b)T}}
\newcommand{\cxjab}{\bar{\bx}_{.j(c)}^{(ab)}}
\newcommand{\cxjabt}{\bar{\bx}_{.j(c)}^{(ab)T}}
\newcommand{\cxjw}{\bar{\bx}_{.j(c)}^{(w)}}
\newcommand{\cxjwt}{\bar{\bx}_{.j(c)}^{(w)T}}
\newcommand{\cyjb}{\bar{y}_{.j(c)}}

\newcommand{\cxijab}{\bx_{ij(c)}^{(ab)}}
\newcommand{\cxijabt}{\bx_{ij(c)}^{(ab)T}}
\newcommand{\cxijw}{\bar{\bx}_{ij(c)}^{(w)}}
\newcommand{\cxijwt}{\bar{\bx}_{ij(c)}^{(w)T}}
\newcommand{\cyijab}{\bar{y}_{ij(c)}}

\newcommand{\cxijkw}{\bx_{ijk(c)}^{(w)}}
\newcommand{\cxijkwt}{\bx_{ijk(c)}^{(w)T}}
\newcommand{\cyijkw}{y_{ijk(c)}}


\section{Notation}\label{notation}

Following \cite{lyu2024increasing}, we arrange the covariate vectors into subvectors corresponding to the different latent random terms in the model because these subvectors produce different asymptotic behaviour.  Specifically,  for model (\ref{two-way cross model no interaction}), 
we arrange the covariate vector $\bx_{ij}$ into three subvectors: a $p_a$-vector of row  covariates $\bxais$, a $p_b$-vector of column covariates $\bxbjs$,  and a $p_w$-vector of within cell covariates $\bxwijs$, where $p=p_a+p_b+p_w \ge p_0$, and then write the regression function in the model (\ref{two-way cross model no interaction}) as
\begin{equation}\label{reg mean fun2}
	\mu(\bx_{ij}) = \xi_0^* + \bxaist\bxi_1^* + \bxbjst\bxi_2^* + \bxwijst\bxi_3^*,
\end{equation}
where $\xi_0^*$ is the intercept, $\bxi_1^*$ is a $p_a$-vector of 
row slopes, $\bxi_2^*$  is a $p_b$-vector of column slopes, and $\bxi_3^*$  is a $p_w$-vector of within-cell slopes. Note that we introduce the asterisk into the notation for covariates and regression parameters in the model with $m=1$ because this reduces possible confusion (over the interaction and within cell terms) with model (\ref{two-way cross model}).  For model (\ref{two-way cross model}) we arrange $\bx_{ijk}$ in four subvectors: a $p_a$-vector of row  covariates $\bxai$, a $p_b$-vector of column covariates $\bxbj$, a $p_{ab}$-vector of interaction covariates $\bxabij$, and a $p_w$-vector of within cell covariates $\bxwijk$, where $p=p_a+p_b+p_{ab}+p_w \ge p_0$, and then set
\begin{equation}\label{reg mean fun}
	\mu(\bx_{ijk})=\xi_0+\bxait\bxi_1+\bxbjt\bxi_2+\bxabijt\bxi_3+\bxwijkt\bxi_4,
 \end{equation}
where $\xi_0$ is the  intercept, $\bxi_1$ is the $p_a$-vector of 
row slopes, $\bxi_2$  is the $p_b$-vector of column slopes, $\bxi_3$  is the $p_{ab}$-vector of interaction slopes, and $\bxi_4$  is the $p_w$-vector of within-cell slopes.  Not all of the vectors need be present in the model; any combination of the types of covariates can be included.

We use a bar above the covariate, and a dot as a placeholder subscript, to represent averages computed by summing the variables across the relevant dimensions (rows, columns, or cells). For example, \(\bar{x}_{i.} = (hm)^{-1} \sum_{j=1}^{h}\sum_{k=1}^{m} x_{ijk}\), and \(\bar{x}_{.j} = (gm)^{-1} \sum_{i=1}^{g}\sum_{k=1}^{m} x_{ijk}\). However, we omit a dot as a placeholder subscript when there is no ambiguity, such as in \(\bar{x}_{ij} = m^{-1}\sum_{k=1}^{m} x_{ijk}\) and \(\bar{x}= n^{-1}\sum_{i=1}^g\sum_{j=1}^h \sum_{k=1}^m x_{ijk}\).
We define averages over the following variables in a similar way: $y_{ijk}$, $e_{ijk}$, $\bxwijk$, $\gamma_{ij}$,  $\bxabij$, $\bxwijs$,$\alpha_i$, $\beta_j$, $\bxai$, $\bxais$, $\bxbj$, $\bxbjs$.
We use the additional subscript $(c)$ for variables that are centered;  see Table~\ref{tab:my_label2}. 
 \begin{table}[!h]
 	\centering
   	\caption{Notation for variables}
 	\label{tab:my_label2}
 	\resizebox{1\linewidth}{!}{
 		\begin{tabular}{lll}
 			\toprule
 		Variable	&Model with no interaction  (\ref{two-way cross model no interaction}) and (\ref{reg mean fun2})  &  Model with interaction (\ref{two-way cross model}) and (\ref{reg mean fun}) \\\hline
 		
   {Row:}  
 		& $ \cxias = \bxais-\barbxas$, $\cxiws=\barbxiws-\barbxws$& 
 			$ \mathbf{x}_{i(c)}^{(a)} = \mathbf{x}_{i}^{(a)}-\bar{\mathbf{x}}^{(a)}$, 
 			$ \bar{\mathbf{x}}_{i.(c)}^{(ab)} = \bar{\mathbf{x}}_{i.}^{(ab)}-\bar{\mathbf{x}}^{(ab)}$  
 	\\
  &		& $ \bar{\mathbf{x}}_{i.(c)}^{(w)} = \bar{\mathbf{x}}_{i.}^{(w)}-\bar{\mathbf{x}}^{(w)}$
 			\\
 			&&
 			\\ \hline
    {Column:}&$\cxjbs=\bxbjs-\barbxbs$, $\cxjws=\barbxjws-\barbxws$&
 			$ \mathbf{x}_{j(c)}^{(b)} = \mathbf{x}_{j}^{(b)}-\bar{\mathbf{x}}^{(b)}$, 
 			$ \bar{\mathbf{x}}_{.j(c)}^{(ab)} = \bar{\mathbf{x}}_{.j}^{(ab)}-\bar{\mathbf{x}}^{(ab)}$
    \\
    &&
			$ \bar{\mathbf{x}}_{.j(c)}^{(w)} = \bar{\mathbf{x}}_{.j}^{(w)}-\bar{\mathbf{x}}^{(w)}$
 			\\
 						&&
 			\\ \hline
    {Interaction:} & & 
 			$ {\mathbf{x}}_{ij(c)}^{(ab)}= {\mathbf{x}}_{ij}^{(ab)}-\bar{\mathbf{x}}_{i.}^{(ab)}-\bar{\mathbf{x}}_{.j}^{(ab)}+\bar{\mathbf{x}}^{(ab)}$\\
    &&
	$ \bar{\mathbf{x}}_{ij(c)}^{(w)}= \bar{\mathbf{x}}_{ij}^{(w)}-\bar{\mathbf{x}}_{i.}^{(w)}-\bar{\mathbf{x}}_{.j}^{(w)}+\bar{\mathbf{x}}^{(w)}$
\\ 
 			 			&&
 			\\\hline
 			Within-cell:  & $\mathbf{x}_{ij(c)}^{*(w)}=\bxwijs-\barbxiws-\barbxjws+\barbxw$&
 			 $ \mathbf{x}_{ijk(c)}^{(w)}=\mathbf{x}_{ijk}^{(w)}-\bar{\mathbf{x}}_{ij}^{(w)}$
 			\\ &&\\
    \bottomrule
 	\end{tabular}}
 \end{table}

The parameters for both models are listed in Table~\ref{tab:my_label}. True parameters are represented by a dot above the symbol (e.g., $\dot\bomega^* = [\dot\xi_0^*, \dot\bxi_1^{*T}, \dot\sigma_\alpha^2, \dot\bxi_2^{T}, \dot\sigma_\beta^2, \dot\bxi_3^{T}, \dot\sigma_e^2]^T$). 
The full mean models include all slope parameters  in (\ref{reg mean fun2}) or (\ref{reg mean fun}). Specific cases where any of the parameters 
are zero are special instances within this framework.

\begin{table}[]
    \centering
        \caption{Notation for parameters}
    \label{tab:my_label}
    \resizebox{1\linewidth}{!}{
    \begin{tabular}{l|ll}
    \toprule
&Model with no interaction  (\ref{two-way cross model no interaction}) and (\ref{reg mean fun2})  &  Model with interaction (\ref{two-way cross model}) and (\ref{reg mean fun}) \\\hline
Parameters: &   $\bomega^*=[\xi_0^*,\bxi_1^{*T},\sigma_\alpha^2, \bxi_2^{*T},\sigma_\beta^2,\bxi_3^{*T},\sigma_e^2]^T$     & 
$\bomega=[\xi_0,\bxi_1^T,\sigma_\alpha^2, \bxi_2^T,\sigma_\beta^2,\bxi_3^T,\sigma_\gamma^2, \bxi_4^T,\sigma_e^2]^T$\\
Fixed effect parameters:&$\bxi^*=[\xi_0^*,\bxi_1^{*T},\bxi_2^{*T},\bxi_3^{*T}]^T$&$\bxi=[\xi_0,\bxi_1^T,\bxi_2^T,\bxi_3^T,\bxi_4^T]^T$ \\
Variance Components:&$\btheta^* = [\siga^2, \sigb^2, \sige^2]^T$ & $\btheta = [\siga^2, \sigb^2, \siggama^2, \sige^2]^T$\\
Intercept:  &$\xi_0^*$ &$\xi_0$\\
Row parameters:& $ \bxi_1^{*},\siga^2$& $  \bxi_1, \siga^2$\\
Column parameters: &$\bxi^{*}_2,\sigb^2$&$\bxi_2,\sigb^2$\\
Interaction parameters: & & $\bxi_3,\siggama^2$\\
Within-cell parameters: &$\bxi^*_3,\sige^2$&$\bxi_4,\sige^2$
\\ \bottomrule
\multicolumn{3}{l}{\textbf{Note:} The intercept  belongs to a row or column parameter depending on the convergence of $g/h$.}
    \end{tabular}}
\end{table}

Let $\hat\bomega^*$ and $\hat\bomega$ be the (working normal) maximum likelihood estimators of $\dot\bomega^*$ and $\dot\bomega$, respectively, obtained by maximizing the normal log-likelihood. The (working normal) REML estimators of $\dot\btheta^*$ and $\dot\btheta$, denoted by $\hat\btheta_R^*$ and $\hat\btheta_R$, are obtained by maximizing the REML criterion. The estimators of $\dot\bxi^*$ and $\dot\bxi$, denoted $\hat\bxi^*(\btheta^*)$ and $\hat\bxi(\btheta)$, are found by solving the maximum likelihood estimating equations with $\btheta^*$ and $\btheta$ fixed. The REML estimators of $\dot\bxi^*$ and $\dot\bxi$ are $\hat\bxi_R^* = \hat\bxi^*(\hat\btheta_R^*)$ and $\hat\bxi_R = \hat\bxi(\hat\btheta_R)$, while the REML estimators of $\dot\bomega^*$ and $\dot\bomega$ are $\hat\bomega_R^*$ and $\hat\bomega_R$.

\section{The two-way crossed design}\label{Sec3: Non Interaction Model}

We derive the asymptotic properties of EBLUPs for individual and cell-level random effects in the model (\ref{two-way cross model no interaction}) and (\ref{reg mean fun2}) under the following condition.

 \medskip
	\noindent
	\textbf{Condition A}
	\begin{enumerate}
  		\item  The model  (\ref{two-way cross model no interaction}) and (\ref{reg mean fun2}) holds with the true parameter $\dot\bomega^*$ an interior point of the parameter space $\bOmega^*$.
		\item  {The number of levels of factor A (rows) $g \to \infty$, and the number of levels of factor B (columns) $h \to \infty$ such that  \(\lim_{g,h\rightarrow\infty} g/h =\eta\) with $\eta \in [0,\infty]$.}
		\item The random variables $\alpha_i$, $\beta_j$ and $e_{ij}$ are independent,  identically distributed and mutually independent.  Moreover, there is a $\delta>0$, such that $\E |\alpha_1|^{4+\delta}<\infty$, $\E |\beta_1|^{4+\delta}<\infty$, and $\E| e_{11}|^{4+\delta}<\infty$.
		\item The limits $ \lim_{g\to\infty} g^{-1} \sumig \bxais$ and $\lim_{h\to\infty} \sumjh \bxbjs$ exist. Let $\bX_{i(c)}^{*(a)}=[\cxiast,\cxiwst]^T$ and $\bX_{j(c)}^{*(b)}=[\cxjbst,\cxjwst]^T$.  Then the limits of the matrices $\lim_{g\to\infty}g^{-1}\sumig\bX_{i(c)}^{*(a)}\bX_{i(c)}^{*(a)T}$, 
  $\lim_{h\to\infty}h^{-1}\sumjh\bX_{j(c)}^{*(b)}\bX_{j(c)}^{*(b)T}$,
  and 
  $\lim_{g,h\to\infty} (gh)^{-1}\sumig\sumjh\bx_{ij(c)}^{*(w)}\bx_{ij(c)}^{*(w)T}$ exist and are positive definite. Additionally,  there is a $\delta>0$, such that 
  \newline
  $\lim_{g\to\infty}g^{-1}\sumig|\cxias|^{2+\delta}<\infty$,
  $\lim_{h\to\infty}h^{-1}\sumjh |\cxjbs|^{2+\delta}<\infty$ 
\\  and $ \lim_{g,h\to\infty}(gh)^{-1}\sumig\sumjh|\bx_{ij(c)}^{*(w)}|^{2+\delta}<\infty$.
	\end{enumerate}
Condition A2 requires $g$ and $h$ to increase to infinity, but it does not constrain the relative rate. We can think of $g$ and $h$ being possibly different increasing functions of a common variable that increases to infinity. Conditions A3 and A4 ensure the existence of the asymptotic variance of the estimating function and that we can establish a Lyapunov condition (and hence a central limit theorem) for the estimating function. They also guarantee convergence of the appropriately normalized derivative of the estimating function. Condition A4 allows us to define the block matrices $\hat\bD_{1}^*=g^{-1}\sum_{i=1}^g \cxias\cxiast$, $\hat\bD_2^*=h^{-1}\sum_{j=1}^h \cxjbs\cxjbst$, and $\hat\bD_3^*=(gh)^{-1}\sum_{i=1}^g\sum_{j=1}^h \bx_{ij(c)}^{*(w)}\bx_{ij(c)}^{*(w)T}$,  and their respective limits  $\bD_{1}^*=\lim_{g\to\infty}\hat\bD_{1}^*$, $\bD_2^*=\lim_{h\to\infty}\hat\bD_{2}^*$, and $\bD_3^*=\lim_{g,h\to\infty}\hat\bD_{3}^*$.

The following theorem for the normal maximum likelihood and normal REML estimators in the model (\ref{two-way cross model no interaction}) and (\ref{reg mean fun2}) can be established from Theorem S.1 of \cite{lyu2024increasing} by multiplying out $\bB^{*-1}\bK^{*-1/2}\bphi^*$ (their notation) in their linear approximation.
\begin{theoremm}\label{theorem2}
	Suppose Condition A holds. 	Then there exist solutions $\hat\bomega^*$ and $\hat\bomega_R^*$ to the normal maximum likelihood estimating equations and to the normal REML estimating equations respectively, such that both estimators have the same linear approximations 
  		\begin{align*}
				\hat\xi_0^*-\dot\xi_0^*&=\frac{1}{g}\sumig(1-\barbxast\bD_1^{*-1}\cxias)\alpha_i+\frac{1}{h}\sumjh(1-\barbxbst\bD_2^{*-1}\cxjbs)\beta_j +o_p(g^{-1/2}+h^{-1/2}),\\
    				\hat{\bxi}_1^*-\dot{\bxi}_1^*&=\bD_{1}^{*-1}\frac{1}{g}\sumig\cxias\alpha_i+ o_p(g^{-1/2}),
    	\qquad
				\hsiga^{*2}-\dsiga^2=\frac{1}{g}\sumig(\alpha_i^2-\dsiga^2) + o_p(g^{-1/2}),
				\\
				\hat{\bxi}_2^*-\dot{\bxi}_2^*&=\bD_2^{*-1}\frac{1}{h}\sumjh\cxjbs\beta_j+ o_p(h^{-1/2}),
				\qquad
				\hsigb^{*2}-\dsigb^2=\frac{1}{h}\sumjh(\beta_j^2-\dsigb^2)+ o_p(h^{-1/2}),
                \\
				\hat{\bxi}_3^*-\dot{\bxi}_3^*&=\bD_3^{*-1}\frac{1}{gh}\sumig\sumjh\cxijws  e_{ij}+ o_p((gh)^{-1/2}),\,\,
			\hsige^{*2}-\dsige^2=\frac{1}{gh}\sumig\sumjh(e_{ij}^{2}-\dsige^2)+ o_p((gh)^{-1/2}).
			\end{align*}
 {When $\eta=0$, we have $\hat\xi_0^*-\dot\xi_0^*=g^{-1}\sumig(1-\barbxast\bD_1^{*-1}\cxias)\alpha_i$; when $\eta=\infty$, we have $\hat\xi_0^*-\dot\xi_0^*=h^{-1}\sumjh(1-\barbxbst\bD_2^{*-1}\cxjbs)\beta_j$.}
 \end{theoremm}

Using Theorem~\ref{theorem2}, we can study the asymptotic distributions of the EBLUPs $\hat{\alpha}_i^*$ and $\hat{\beta}_j^*$, defined  either as  $\hat{\alpha}_i^*(\hat{\bomega}^*)$, $\hat{\beta}_j^*(\hat{\bomega}^*)$ or $\hat{\alpha}_i^*(\hat{\bomega}_R^*)$, $\hat{\beta}_j^*(\hat{\bomega}_R^*)$, where
\begin{equation}\label{blups2}
\begin{split}
\hat\alpha_i^*(\bomega^*)&=\frac{h\siga^2}{\lambda_2^*}\{\yidd-\frac{1}{h}\sumjh\mu(\bx_{ij})\}-h\siga^2(\frac{1}{\lambda_2^*}-\frac{1}{\lambda_4^*})\{\yddd-\frac{1}{gh}\sumig\sumjh\mu(\bx_{ij})\},
\\
\hat\beta_j^*(\bomega^*)&=\frac{g\sigb^2}{\lambda_3^*}\{\ydjd-\frac{1}{g}\sumig\mu(\bx_{ij})\}-g\sigb^2(\frac{1}{\lambda_3^*}-\frac{1}{\lambda_4^*})\{\yddd-\frac{1}{gh}\sumig\sumjh\mu(\bx_{ij})\},
\end{split}
\end{equation}
with $\lambda_1^*=\sige^2$, $\lambda_2^*=\sige^2+h\siga^2$, $\lambda_3^*=\sige^2+g\sigb^2$ and $\lambda_4^*=\sige^2+h\siga^2+g\sigb^2$.
Since the results are the same for both estimators, we make no distinction 
between them in what follows.   The proofs are available in the Supplementary Material. We begin by establishing an asymptotic linearity result for the EBLUPs in model~(\ref{two-way cross model no interaction}).

\begin{theoremm}\label{theorem4}
	Suppose Condition A holds. Define  $H_{is}^{*(a)}=1+\cxiast\bD_1^{*-1}\boldsymbol{x}_{s(c)}^{*(a)}$ and $H_{jt}^{*(b)}=1+\cxjbst\bD_2^{*-1}\boldsymbol{x}_{t(c)}^{*(b)}$.  Then we have
	\begin{align*}
		&\hat{\alpha}_i^*-\alpha_i=\eidd-\frac{1}{g}\sum_{s=1}^gH_{is}^{*(a)}\alpha_s
		+o_p(g^{-1/2}+ h^{-1/2}),
		\\&
		\hat\beta_j^*-\beta_j=\edjd-\frac{1}{h}\sum_{t=1}^h H_{jt}^{*(b)}\beta_t
		+o_p(g^{-1/2}+ h^{-1/2}).
	\end{align*}
\end{theoremm}

The asymptotic distribution of the EBLUPs in model~(\ref{two-way cross model no interaction}) is then given in the following corollary.
\begin{corollary}\label{corollary2}
    Suppose Condition A holds, $i \ne i'$ and $j \ne j'$, and write $\hat{\boldsymbol{\chi}}^*=[\hat{\alpha}_i^*-\alpha_i,\,
    \hat{\alpha}_{i'}^*-\alpha_{i'},\,
    \hat{\beta}_j^*-\beta_j,\,
    \hat{\beta}_{j'}^*-\beta_{j'}]^T$. Then as $g, h\to\infty$ such that $\lim_{g,h \rightarrow \infty} g/h = \eta$, we have
for $0 \le \eta < \infty$,
\begin{align*}
g^{1/2}\hat{\boldsymbol{\chi}}^*\xrightarrow{D}  N\left(\boldmath{0},
\diag(\boldsymbol{F}^*, \boldsymbol{G}^*)
\right), \mbox{ where } \boldsymbol{F}^* = \begin{bmatrix}
F_{i,i}^{*} & F_{i,i'}^{*}\\
 F_{i',i}^{*}&F_{i',i'}^{*} 
\end{bmatrix},
\boldsymbol{G}^* = \begin{bmatrix}
G_{j,j}^{*} & G_{j,j'}^{*}\\
 G_{j',j}^{*}&G_{j',j'}^{*} 
\end{bmatrix},
\end{align*}
with $F_{su}^* = {\eta\dsige^2}I(s=u) + {\dsiga^2}H_{su}^{*(a)}$ and $G_{tv}^*={\dsige^2}I(t=v)+ {\eta\dsigb^2}H_{tv}^{*(b)}$.
For $0 < \eta \le \infty$, we can normalise by $h^{1/2}$ instead of $g^{1/2}$; the asymptotic covariance matrix is $\eta^{-1}\diag(
\boldsymbol{F}^*, \boldsymbol{G}^*)$.
\end{corollary}

\noindent
We can also use different normalisation for the different random effects; we omit the formal statement.
Finally, we can derive the asymptotic distribution of the cell-level EBLUPs.
\begin{corollary}\label{corollary3}
    Suppose Condition A and at least one of $i\ne i'$ or $j\ne j'$ holds. Then as $g, h\to\infty$ such that $\lim_{g,h \rightarrow \infty} g/h = \eta$, we have
for $0 \le \eta < \infty$,
\begin{align*}
&g^{1/2}\begin{pmatrix}
     \hat{\alpha}_i^*+ \hat{\beta}_j^*-\alpha_{i}-\beta_{j}\\
          \hat{\alpha}_{i'}^*+ \hat{\beta}_{j'}^*-\alpha_{i'}-\beta_{j'}
\end{pmatrix}
\xrightarrow{D}  N\left(\begin{bmatrix}
    0\\0
\end{bmatrix},
\begin{bmatrix}
C_{ij,ij}^{*} 
& C_{ij,i'j'}^{*}\\
 C_{ij,i'j'}^{*}&
C_{i'j',i'j'}^{*} 
\end{bmatrix}
\right),
\end{align*}
where $C_{st,uv}^{*}= \dsige^2\{\eta I(s=u) + I(t=v)\}+{\dsiga^2}H_{su}^{*(a)}+{\eta\dsigb^2} H_{tv}^{*(b)}$.
For $0<\eta\le \infty$, we can normalise by $h^{1/2}$ instead of $g^{1/2}$; the elements of the asymptotic covariance matrix are $\eta^{-1}C_{st,uv}^{*}$.   
In both cases, the cell level EBLUPs are asymptotically dependent even for cells in different rows and/or columns ($i\ne i'$ and $j\ne j'$). 
\end{corollary}
\noindent  
\textbf{Remark 1.} {Practical approximations to the asymptotic distributions of the EBLUPs can be obtained by setting $\eta=g/h$ in both Corollaries~\ref{corollary2} and \ref{corollary3}. The various normalisations lead to the same approximations.}
We can use these approximations to further approximate the distributions of the EBLUPs by convolutions of their true distributions and normal distributions (Details can be found in the Supplementary Material); see \cite{lyu2019eblup} for analogous approximations for the nested case. 
\section{The replicated two-way crossed design}\label{Sec4: Interaction Model}

We derive the asymptotic properties of the EBLUPs for individual and cell-level random effects in the  model (\ref{two-way cross model}) and (\ref{reg mean fun}) under the following condition.

 \medskip
	\noindent
	\textbf{Condition B}
	\begin{enumerate}
		\item  The model (\ref{two-way cross model}) and (\ref{reg mean fun}) holds with the true parameter $\dot\bomega$ an interior point of the parameter space $\bOmega$.
		\item  {The number of levels of factor A (rows) $g \to \infty$, the number of levels of factor B (columns) $h \to \infty$ such that  \(\lim_{g,h\rightarrow\infty} g/h =\eta\) with $\eta \in [0,\infty]$, and 
        the number of observations within each cell $m \to \infty$ such that \(\lim_{g,h\rightarrow\infty} g/m =\eta_1\) and \(\lim_{g,h\rightarrow\infty} h/m =\eta_2\) with $\eta_1, \eta_2 \in [0,\infty]$.}
		\item The random variables $\alpha_i$, $\beta_j$, $\gamma_{ij}$ and $e_{ijk}$ are independent, identically distributed and mutually independent.  Moreover, there is a $\delta>0$, such that $\E |\alpha_1|^{4+\delta}<\infty$, $\E |\beta_1|^{4+\delta}<\infty$, $\E |\gamma_{11}|^{4+\delta}<\infty$ and $\E| e_{111}|^{4+\delta}<\infty$.
		\item Let $\bX_{i(c)}^{(a)}=[\cxiat,\cxiabt,\cxiwt]^T$,
$\bX_{j(c)}^{(b)}=[\cxjbt,\cxjabt,\cxjwt]^T$ and 
$\bX_{ij(c)}^{(ab)}=[\cxijabt,\cxijwt]^T$.  The limits $\lim_{g\to\infty} g^{-1} \sumig \bxai$ and $\lim_{h\to\infty} \sumjh \bxbj$ exist. Also, the limits of the matrices $\lim_{g\to\infty}g^{-1}\sumig\bX_{i(c)}^{(a)}\bX_{i(c)}^{(a)T}$, 
  $\lim_{h\to\infty}h^{-1}\sumjh\bX_{j(c)}^{(b)}\bX_{j(c)}^{(b)T}$, \\$\lim_{g,h\to\infty} (gh)^{-1}\sumig\sumjh\bX_{ij(c)}^{(ab)}\bX_{ij(c)}^{(ab)T}$ and 
  $\lim_{g,h,m\to\infty} n^{-1}\sumig\sumjh\sumkm\cxijkw\cxijkwt$ exist and are positive definite. Additionally,  there is a $\delta>0$, such that\\  $\lim_{g\to\infty}g^{-1}\sumig|\cxia|^{2+\delta}<\infty$, 
  $\lim_{h\to\infty}h^{-1}\sumjh |\cxjb|^{2+\delta}<\infty$,\\ $\lim_{g,h\to\infty}(gh)^{-1}\sumig\sumjh|\cxijab|^{2+\delta}<\infty$, and\\  $ \lim_{g,h,m\to\infty}n^{-1}\sumig\sumjh\sumkm|\cxijkw|^{2+\delta}<\infty$.
	\end{enumerate}
Condition B is similar to Condition A. Again, we can think of $g$, $h$ and $m$ being possibly different increasing functions of a common variable that increases to infinity. Condition B2 allows $g$, $h$, $m$ to diverge to infinity without imposing constraints on the relative rates.  In practice, we expect $m$ to grow more slowly than $g$ and $h$ so that $\eta_1=\eta_2 = \infty$ is more important than $\eta_1=\eta_2=0$.  When $\eta_1, \eta_2 \in (0,\infty)$, one of $\eta_1$ or $\eta_2$ is redundant, but we retain both for generality.
	Conditions B3 and B4 ensure that limits needed for the existence of the asymptotic variance of the estimating function exist and that we can establish a Lyapunov condition (and hence a central limit theorem) for the estimating function. They also ensure convergence of the appropriately normalized second derivative of the estimating function.  Condition B4 allows us to 
    define the block matrices $\hat\bD_{1}=g^{-1}\sumig\cxia\cxiat$, $\hat\bD_2=h^{-1}\sumjh\cxjb\cxjbt$, $\hat\bD_3= (gh)^{-1}\sum_{i,j}\cxijab\cxijabt$ and $\hat\bD_4=n^{-1}\sum_{i,j,m}\cxijkw\cxijkwt$,  and their respective limits  $\bD_{1}=\lim_{g\to\infty}\hat\bD_{1}$, $\bD_2=\lim_{h\to\infty}\hat\bD_{2}$, $\bD_3=\lim_{g,h\to\infty}\hat\bD_{3}$ and $\bD_4=\lim_{g,h,m\to\infty}\hat\bD_{4}$. 
\color{black}

The following theorem for the normal maximum likelihood and normal REML estimators in the model (\ref{two-way cross model}) and (\ref{reg mean fun}) can be established from Theorem 1 of \cite{lyu2024increasing} by multiplying out $\bB^{-1}\bK^{-1/2}\bphi$ (their notation) in their linear approximations.
	\begin{theoremm}\label{theorem1}
    Suppose Condition B holds. Then there exist solutions $\hat\bomega$ and $\hat\bomega_R$ to the normal maximum likelihood estimating equations and   to the normal REML estimating equations respectively, such that both estimators have the same linear approximations
  		\begin{align*}
				\hat\xi_0-\dot\xi_0&=\frac{1}{g}\sumig(1-\barbxat\bD_1^{-1}\cxia)\alpha_i+\frac{1}{h}\sumjh(1-\barbxbt\bD_2^{-1}\cxjb)\beta_j,\\
    				\hat{\bxi}_1-\dot{\bxi}_1&=\bD_{1}^{-1}\frac{1}{g}\sumig\cxia\alpha_i,
    	\qquad
				\hsiga^2-\dsiga^2=\frac{1}{g}\sumig(\alpha_i^2-\dsiga^2),
				\\
				\hat{\bxi}_2-\dot{\bxi}_2&=\bD_2^{-1}\frac{1}{h}\sumjh\cxjb\beta_j,
				\qquad
				\hsigb^2-\dsigb^2=\frac{1}{h}\sumjh(\beta_j^2-\dsigb^2),
                \\
				\hat{\bxi}_3-\dot{\bxi}_3&=\bD_{3}^{-1}\frac{1}{gh}\sumig\sumjh\cxijab\gamma_{ij},\qquad
				\hsiggama^2-\dsiggama^2=\frac{1}{gh}\sumig\sumjh(\gamma_{ij}^2-\dot{\sigma}_{\gamma}^2),\\
				\hat{\bxi}_4-\dot{\bxi}_4&=\bD_4^{-1}\frac{1}{n}\sumig\sumjh\sumkm\cxijkw  e_{ijk},\qquad
			\hsige^2-\dsige^2=\frac{1}{n}\sumig\sumjh\sumkm(e_{ijk}^2-\dsige^2).
			\end{align*}
{When $\eta=0$, we have $\hat\xi_0-\dot\xi_0=g^{-1}\sumig(1-\barbxat\bD_1^{-1}\cxia)\alpha_i$; when $\eta=\infty$, we have $\hat\xi_0-\dot\xi_0=h^{-1}\sumjh(1-\barbxbt\bD_2^{-1}\cxjb)\beta_j$.}
\end{theoremm}
   



Analogously to the previous section, we can use Theorem~\ref{theorem1} to study the asymptotic distribution of the EBLUPs ($\hat\alpha_i, \hat\beta_j$ and $\hat\gamma_{ij}$). 
As before, the results are the same for the maximum likelihood and REML parameter estimators, so we again make no distinction 
between them in what follows. The proofs are available in the Supplementary Material.
We begin by establishing an asymptotic linearity result for EBLUPs in model~(\ref{two-way cross model}).

\begin{theoremm}\label{theorem3}
   Suppose Condition B holds.  We define $H_{is}^{(a)}=1+\cxiat\bD_1^{-1}\boldsymbol{x}^{(a)}_{s(c)}$ and $H_{jt}^{(b)}=1+\cxjbt\bD_2^{-1}\boldsymbol{x}^{(b)}_{t(c)}$.  Then we have
   \begin{align*}
       &\hat{\alpha}_i-\alpha_i=\gammaid-\frac{1}{g}\sum_{s=1}^gH_{is}^{(a)} \alpha_s+o_p(g^{-1/2}+ h^{-1/2}),
\\&
\hat\beta_j-\beta_j=\gammadj-\frac{1}{h}\sum_{t=1}^h H_{jt}^{(b)}\beta_t
+o_p(g^{-1/2}+ h^{-1/2}),\\&
\hat\gamma_{ij}-\gamma_{ij}
=\eijd-\gammaid-\gammadj+O_p(m^{-1})+o_p(g^{-1/2}+ h^{-1/2}).
   \end{align*}
\end{theoremm}
The asymptotic distribution of the EBLUPs in model~(\ref{two-way cross model}) is then given in the following corollary.
\begin{corollary}\label{corollary1}
    Suppose Condition B, $i\ne i'$ and $j\ne j'$ hold. Write $\hat{\boldsymbol{\chi}} = [\hat{\alpha}_i - \alpha_i,\,\hat{\alpha}_{i'} - \alpha_{i'},\, \hat{\beta}_j - \beta_j,\,\hat{\beta}_{j'} - \beta_{j'},\, \hat{\gamma}_{ij} - \gamma_{ij},\, \hat{\gamma}_{i'j} - \gamma_{i'j},\, \hat{\gamma}_{ij'} - \gamma_{ij'},\, \hat{\gamma}_{i'j'} - \gamma_{i'j'}]^T$ and define $\eta=\lim_{g,h \to\infty}g/h$, $\eta_1=\lim_{g,m \to\infty}g/m$, and $\eta_2=\lim_{h,m \to\infty} h/m$. Then as $g, h , m\to\infty$, we have  for $0 \le \eta,\eta_1 < \infty$,
\begin{align*}
g^{1/2} \hat{\boldsymbol{\chi}}\xrightarrow{D}  N\left(\boldmath{0},
\begin{bmatrix} \diag(\boldsymbol{F}, \boldsymbol{G}) &\boldsymbol{U} \\
\boldsymbol{U}^T & \eta_1 \dsige^2 \boldsymbol{I}_4 + \boldsymbol{Q}
\end{bmatrix}
\right), \mbox{ where } \boldsymbol{F} = \begin{bmatrix}
F_{i,i} & F_{i,i'}\\
 F_{i',i}&F_{i',i'} 
\end{bmatrix},
\boldsymbol{G} = \begin{bmatrix}
G_{j,j} & G_{j,j'}\\
 G_{j',j}&G_{j',j'} 
\end{bmatrix},
\end{align*}
with $F_{su} = {\eta\dsiggama^2}I(s=u) + {\dsiga^2}H_{su}^{(a)}$ and $G_{tv}={\dsiggama^2}I(t=v)+ {\eta\dsigb^2}H_{tv}^{(b)}$, and
\begin{align*}
\boldsymbol{U} = -\dsiggama^2 \begin{bmatrix}
\eta& 0 & \eta & 0\\
0 & \eta & 0 & \eta\\
1 & 1 &  0 & 0\\
0 & 0 & 1 & 1
\end{bmatrix},
\quad
\boldsymbol{Q} = \dsiggama^2 \begin{bmatrix}
1+\eta& 1 & \eta & 0\\
1 & 1+\eta & 0 & \eta\\
\eta & 0 &  1+\eta & 1\\
0 & \eta & 1 & 1+\eta
\end{bmatrix}.
\end{align*}  
For $0< \eta \le \infty$ and $0\le \eta_2<\infty$, we can normalise by $h^{1/2}$ instead of $g^{1/2}$;  the asymptotic covariance matrix has $\eta_2$ in place of $\eta_1$ and each of the matrices $\boldsymbol{F}$, $\boldsymbol{G}$, $\boldsymbol{U}$, and $\boldsymbol{Q}$ multiplied by $\eta^{-1}$. For $0 < \eta_1, \eta_2 \le \infty$, we can normalise by $m^{1/2}$ instead of $g^{1/2}$; the asymptotic covariance matrix has $F_{su} = {\eta_2^{-1}\dsiggama^2}I(s=u) + \eta_1^{-1}{\dsiga^2}H_{su}^{(a)}$ and $G_{tv}=\eta_1^{-1}{\dsiggama^2}I(t=v)+ {\eta_2^{-2}\dsigb^2}H_{tv}^{(b)}$, $\eta_1$ replaced by $1$ and
\begin{align*}
\boldsymbol{U} = -\dsiggama^2 \begin{bmatrix}
\eta_2^{-1}& 0 & \eta_2^{-1} & 0\\
0 & \eta_2^{-1} & 0 & \eta_2^{-1}\\
\eta_1^{-1} & \eta_1^{-1} &  0 & 0\\
0 & 0 & \eta_1^{-1} & \eta_1^{-1}
\end{bmatrix},
\boldsymbol{Q} = \dsiggama^2 \begin{bmatrix}
\eta_1^{-1}+\eta_2^{-1}& \eta_1^{-1} & \eta_2^{-1} & 0\\
\eta_1^{-1} & \eta_1^{-1}+\eta_2^{-1} & 0 & \eta_2^{-1}\\
\eta_2^{-1} & 0 &  \eta_1^{-1}+\eta_2^{-1} & \eta_1^{-1}\\
0 & \eta_2^{-1} & \eta_1^{-1} & \eta_1^{-1}+\eta_2^{-1}
\end{bmatrix}.
\end{align*}   
\end{corollary}


\noindent We can again consider different normalisation for the different random effects.  Our final result is for the cell-level EBLUPs.
\begin{corollary}\label{corollary8}
    Suppose Condition B and at least one of $i\ne i'$ or $j\ne j'$ holds. Then as $g, h\to\infty$ such that $\lim_{g,h \rightarrow \infty} g/h = \eta$, $\lim_{g,m \rightarrow \infty} g/m = \eta_1$ and $\lim_{h,m \rightarrow \infty} h/m = \eta_2$, we have
for $0 < \eta, \eta_1 \le \infty$,
\begin{align*}
&g^{1/2}\begin{pmatrix}
     \hat{\alpha}_i+ \hat{\beta}_j + \hat{\gamma}_{ij}-\alpha_{i}-\beta_{j} - \gamma_{ij}\\
          \hat{\alpha}_{i'}+ \hat{\beta}_{j'} + \hat{\gamma}_{i'j'}-\alpha_{i'}-\beta_{j'} - \gamma_{i'j'}
\end{pmatrix}
\xrightarrow{D}  N\left(\begin{bmatrix}
    0\\0
\end{bmatrix}, \eta_1\dsige^2\boldsymbol{I}_2 + 
\begin{bmatrix}
C_{ij,ij} 
& C_{ij,i'j'}\\
 C_{ij,i'j'}&
C_{i'j',i'j'} 
\end{bmatrix}
\right),
\end{align*}
where $C_{st,uv}= {\dsiga^2}H_{su}^{(a)}+{\eta\dsigb^2} H_{tv}^{(b)}$.  For $0 < \eta \le \infty$ and $0 \le \eta_2 < \infty$, we can normalize by $h^{1/2}$ instead of $g^{1/2}$; the asymptotic covariance matrix has $\eta_2$ in place of $\eta_1$ and $C_{st,uv}$ is multiplied by $\eta^{-1}$.  
For $0< \eta_1, \eta_2 \le \infty$, we can normalize by $m^{1/2}$ instead of $g^{1/2}$; the asymptotic covariance matrix has $1$ in place of $\eta_1$ and $C_{st,uv}= \eta_1^{-1}{\dsiga^2}H_{su}^{(a)}+{\eta_2^{-1}\dsigb^2} H_{tv}^{(b)}$.
The cell level EBLUPs are asymptotically dependent even for cells in different rows and/or columns ($i\ne i'$ and $j\ne j'$). 
\end{corollary}
\noindent
\textbf{Remark 2.}
{Analogously to Remark 1, practical approximations to the asymptotic distributions of the EBLUPs can be obtained by setting $\eta=g/h$, $\eta_1=g/m$ and $\eta_2=h/m$ in both Corollaries~\ref{corollary1} and \ref{corollary8}; the different normalisations lead to the same approximations; and we can use these approximations to further approximate the distributions of the EBLUPs by convolutions of their true distributions and normal distributions  (Details can be found in the Supplementary Material). }
\section{Approximate Mean Squared Errors of EBLUPs}\label{Sec4:MSE}

Corollaries~\ref{corollary2} and~\ref{corollary1} provide very simple and interpretable approximations to the asymptotic mean squared error (as $g, h,m\to\infty$) of the EBLUPs, namely,
\begin{align}
&\operatorname{MSE}_{LSW}(\hat\alpha_i^*)=\frac{\dsige^2}{h}+\frac{\dsiga^2}{g} \hat H_{ii}^{*(a)},
&& \operatorname{MSE}_{LSW}(\hat\beta_j^*)=\frac{\dsige^2}{g}+\frac{\dsigb^2}{h}\hat  H_{jj}^{*(b)},\label{MSE_LSW 1}\\
&\operatorname{MSE}_{LSW}(\hat\alpha_i)=\frac{\dsiggama^2}{h}+\frac{\dsiga^2}{g}\hat H_{ii}^{(a)},
&& \operatorname{MSE}_{LSW}(\hat\beta_j)=\frac{\dsiggama^2}{g}+\frac{\dsigb^2}{h} \hat H_{jj}^{(b)},
\nonumber\\&
 \operatorname{MSE}_{LSW}(\hat\gamma_{ij})=\frac{\dsige^2}{m}+(\frac{1}{g}+\frac{1}{h})\dsiggama^2,\label{MSE_LSW 2}
 \end{align}
where $\hat H_{ii}^{*(a)}=1+\cxiast\hat{\bD}_1^{*-1}\cxias$,
 $\hat H_{jj}^{*(b)}=1+\cxjbst\hat{\bD}_2^{*-1}\cxjbs$, $\hat H_{ii}^{(a)}=1+\cxiat\hat{\bD}_1^{-1}\cxia$, and
 $\hat H_{jj}^{(b)}=1+\cxjbt\hat{\bD}_2^{-1}\cxjb$
 are the sample versions of $ H_{ii}^{*(a)}$, $ H_{jj}^{*(b)}$, $ H_{ii}^{(a)}$ and $ H_{jj}^{(b)}$, $1\le i\le g$, $1\le j\le h$. 
It is straightforward to estimate the asymptotic mean squared error of the EBLUPs by replacing their unknown variance components by the maximum likelihood or REML estimators,
and we can then use these estimators of the asymptotic mean squared error to construct simple asymptotic $100(1-q/2)\%$ prediction intervals for EBLUPs as
\begin{align}
&
\hat\alpha_i^*\pm\mathcal{N}_{q}\widehat{\operatorname{MSE}}_{LSW}^{1/2}(\hat\alpha_i^*),
&&
\hat\beta_j^*\pm\mathcal{N}_{q}\widehat{\operatorname{MSE}}_{LSW}^{1/2}(\hat\beta_j^*),
\label{prediction interval 2}\\
&\hat\alpha_i\pm\mathcal{N}_{q}\widehat{\operatorname{MSE}}_{LSW}^{1/2}(\hat\alpha_i),
&&
\hat\beta_j\pm\mathcal{N}_{q}\widehat{\operatorname{MSE}}_{LSW}^{1/2}(\hat\beta_j),
&&
\hat\gamma_{ij}\pm\mathcal{N}_{bq}\widehat{\operatorname{MSE}}_{LSW}^{1/2}(\hat\gamma_{ij}),
\label{prediction interval 1}
\end{align}
where $\mathcal{N}_{q} =\Phi^{-1}(1-q/2)$ and $\Phi$ denotes the cumulative distribution function of the standard normal distribution.
The asymptotic coverage of these interval is guaranteed by Corollaries~\ref{corollary2}, ~\ref{corollary1} and Slutsky's Theorem, and holds without any assumption of normality in the models.   Corollaries~\ref{corollary3} and~\ref{corollary8} analogously provide approximations to the asymptotic mean squared errors that lead to prediction intervals for the cell level EBLUPS.

\newcommand{\brho}{\boldsymbol{\rho}}

Other approximations for the mean squared error of EBLUPs have been obtained by \cite{kackar1984approximations} and \cite{prasad1990estimation}.  They considered only the nested case, but their approach can be generalized to crossed cases, even though the asymptotic framework and the properties of the estimators are unclear.  Their approaches are based on taking the limit of the mean square error of the EBLUPs rather than the mean squared error of an asymptotically linear approximation, and they assumed that all random terms in the models are normally distributed.

We illustrate the approximations for $\hat{\alpha}(\hat\btheta)$; the approximations for the other EBLUPS are given in the Supplementary Material.  For a model in the general form (\ref{two-way matrix}), 
we can write the BLUP for $\alpha_i$ in matrix form as 
\begin{align*}
&\hat\alpha_i(\btheta)=\siga^2\brho_i^{(a)T}\bZ_1^T\bP(\btheta)\by,
\end{align*}
where $\bP(\btheta)=\bV^{-1}(\btheta)-\bV^{-1}(\btheta)\bX\{\bX^T\bV^{-1}(\btheta)\bX\}^{-1}\bX^T\bV^{-1}(\btheta)$, and  $\brho_i^{(a)}$
is a $g$-vector with all elements zero except for the one at the $i$th, 
element.
For translation invariant estimators $\hat\btheta$ (which includes the maximum likelihood and REML estimators) and under the assumption of normality, 
\cite{kackar1984approximations} and \cite{prasad1990estimation} wrote the mean squared error as 
\begin{displaymath}
\resizebox{\textwidth}{!}{
$\begin{aligned}
\operatorname{MSE}(\hat\alpha_i)=\E\{\hat\alpha_i(\hat\btheta)-\alpha_i\}^2&=\E\{\hat\alpha_i(\dot\btheta)-\alpha_i\}^2+\E\{\hat\alpha_i(\hat\btheta)-\hat\alpha_i(\dot\btheta)\}^2
=M_{1i}(\dot\btheta)+M_{2i}(\dot\btheta), 
\end{aligned}$}
\end{displaymath}
say. It is straightforward to show that 
\begin{align}
&  M_{1i}(\dot\btheta)=\dsiga^2-\dsiga^4 \brho_i^{(a)T}\bZ_1^T\bP(\dot\btheta)\bZ_1\brho_i^{(a)}.\label{MSE:M1}
\end{align}
As $\hat\alpha(\hat\btheta)$ is not linear, usable expressions for $M_{2i}(\dot\btheta)$ can only be obtained by approximation. Making an expansion under the expectation means that additional conditions are required to establish that $\E o_p(g^{-1})=o(g^{-1})$ in order to control the approximation error.
\cite{kackar1984approximations} expanded $\hat\alpha_i(\btheta)$ and approximated the expectation of a product by the product of expectations to derive the approximation $M_{KH,2i}(\dot\btheta)$ to $M_{2i}(\dot\btheta)$, where
\begin{align}
M_{KH,2i}(\btheta)=\tr\{\mathcal{A}_i^{(a)}(\btheta)\mathcal{B}(\btheta)\}, \label{MSE:M2KH}
\end{align}
with $	\mathcal{A}_i^{(a)}(\btheta)=\var\{\bd_i^{(a)}(\btheta)\}$ with $	\bd_i^{(a)}(\btheta)=[{\partial \hat\alpha_i(\btheta)}/{\partial \sige^2},{\partial \hat\alpha_i(\btheta)}/{\partial \siga^2},{\partial \hat\alpha_i(\btheta)}/{\partial \sigb^2},{\partial \hat\alpha_i(\btheta)}/{\partial \sigma_{\gamma}^2}]^T$,
$\mathcal{B}(\btheta)=[B_{st}(\btheta)]$, $B_{st}(\btheta)=2\operatorname{trace}\{\bV^{-1}(\btheta)\bZ_s\bZ_s^T\bV^{-1}(\btheta)\bZ_t\bZ_t^T\}]^{-1}$ for $ s,t=0,1,2,3$, and $\bZ_0=\bI_n$.
\cite{prasad1990estimation} proposed a slightly different expansion, yielding the approximation $M_{PR,2i}(\dot\btheta)$, where
\begin{align}
	M_{PR,2i}(\btheta)=\tr\{\boldsymbol{\varGamma}_i^{(a)}(\btheta) \bV(\btheta)\boldsymbol{\varGamma}^{(a)T}_i(\btheta)\mathcal{B}(\btheta)\}, \label{MSE:M2PR}
\end{align}
with $\boldsymbol{\varGamma}^{(a)}_i(\btheta)=[\varGamma_{i0}^{(a)}(\btheta),\varGamma_{i1}^{(a)}(\btheta),\varGamma_{i2}^{(a)}(\btheta),\varGamma_{i3}^{(a)}(\btheta)]^T$,  with
$\varGamma_{i0}^{(a)}(\btheta)=-\siga^2\brho_i^{(a)T}\bZ_1^T\bV^{-2}(\btheta)$,
$\varGamma_{i1}^{(a)}(\btheta)=\brho_i^{(a)T}\bZ_1^T\bV^{-1}(\btheta)-\siga^2\brho_i^{(a)T}\bZ_1^T\bV^{-1}(\btheta)\bZ_1\bZ_1^T\bV^{-1}(\btheta)$,
\newline
$\varGamma_{i2}^{(a)}(\btheta)=-\siga^2\brho_i^{(a)T}\bZ_1^T\bV^{-1}(\btheta)\bZ_2\bZ_2^T\bV^{-1}(\btheta)$,
and $\varGamma_{i3}^{(a)}(\btheta)=-\siga^2\brho_i^{(a)T}\bZ_1^T\bV^{-1}(\btheta)\bZ_3\bZ_3^T\bV^{-1}(\btheta)$.
%
%
%
Neither approximation simplifies readily, and their application to large datasets can be challenging  due to the number of matrix operations that need to be applied to high dimensional matrices.  This quickly  leads to memory exhaustion in computational environments like R or Python.

\section{Simulations}\label{Sec6:simulation}


For the two-way crossed random effect with no interaction model (\ref{two-way cross model no interaction}), we carried out simulations over a range of settings to compare the finite sample performance of our prediction intervals for random effects (\ref{prediction interval 2}) using our estimated mean squared errors (\ref{MSE_LSW 1}) with prediction intervals of the same form but with  (\ref{MSE_LSW 1}) replaced by the estimators of \cite{kackar1984approximations} and \cite{prasad1990estimation} given by $\widehat{\operatorname{MSE}}_{KH,i}=M_{1i}(\hat{\btheta})+M_{KH,2i}(\hat{\btheta})$ and $\widehat{\operatorname{MSE}}_{PR,i}=M_{1i}(\hat{\btheta})+2M_{PR,2i}(\hat{\btheta})$, respectively, where $M_{1i}(\btheta)$ is defined in (\ref{MSE:M1}), $M_{KH,2i}(\btheta)$ in (\ref{MSE:M2KH}) and $M_{PR,2i}(\btheta)$ in (\ref{MSE:M2PR}).

We generated data for $g \in \{10, 50,100\}$ and $h \in \{10,50,100\}$.  The covariate $x_{ij}$ was generated with a crossed structure by setting $x_{ij} = 4 +t_i+ 1.5u_j + 2v_{ij}$, where $t_i$, $u_j$, and $v_{ij}$ are independent standard normal random variables.  The random effects were generated independently with $\alpha_i$ generated from $F_{\alpha}$, $\beta_j$ from  $F_{\beta}$, and $e_{ij}$ from $F_{e}$.   We set $F_\alpha=N(0,\dsiga^2)$ with $\dsiga^2 =9$,
 $F_\beta=N(0,\dsigb^2)$ or $F_\beta=0.3N(0.5,1)+0.7N(\mu,(\dsigb^2-0.375-0.7\mu^2)/0.7)$ with $\dsigb^2 =49$,
and
 $F_e=N(0,\dsige^2)$ or $F_e=0.3N(0.5,1)+0.7N(\mu,(\dsige^2-0.375-0.7\mu^2)/0.7)$, with $\dsige^2=81$.  Here $\mu=-0.3\times 0.5/0.7$ to make the mean of the mixture distribution zero.
 We then computed the response using the model
\begin{align*}
 y_{ij}=&\xddd\xi_0+(\xidd-\xddd)\xi_1+(\xdjd-\xddd)\xi_2+(x_{ij}-\xidd-\xdjd+\xddd)\xi_3  
 +\alpha_i+\beta_j+e_{ij},
\end{align*}
for $ i=1,\ldots,g,\, j=1,\ldots,h,$ with true parameters $\dot\bxi = [0,5,7,3]^T$. We explored 36 scenarios resulting from the combination of the 9 $(g,h)$ configurations with the 4 additional random effect settings ($F_\beta$ and $F_e$ are normal or mixture distributions). 

For the two-way crossed random effect with interaction model (\ref{two-way cross model}) and (\ref{reg mean fun}), the MSE approximation expressions from \cite{kackar1984approximations} and \cite{prasad1990estimation} become exceedingly complex, even for relatively small sizes (e.g., $g=30, h=30$, and $m=15$, leading to $n=13,500$). This complexity poses significant challenges when computing $\{\bX^T\bV^{-1}(\btheta)\bX\}^{-1}$.  We cannot obtain a simplified exact expression for $\{\bX^T\bV^{-1}(\btheta)\bX\}^{-1}$, and dealing with a $13,500\times 13,500$ matrix in R often exhausts the memory. 
Consequently, our simulations within model (\ref{two-way cross model}) were limited to examining the finite sample performance of the prediction intervals for random effects (\ref{prediction interval 1}) across various settings, utilizing our estimated mean squared error equations (\ref{MSE_LSW 2}), without conducting comparisons with intervals using the \cite{kackar1984approximations} and \cite{prasad1990estimation} approximations.

When including interactions, we generated data  with $g \in \{10, 50\}$, $h \in \{10,50\}$, and $m \in \{10, 30\}$.  The covariate $x_{ijk}$ was again generated with a crossed structure by setting $x_{ijk} = 4 +t_i+ 1.5u_j + 2v_{ij} + 3w_{ijk}$, where $t_i$, $u_j$, $v_{ij}$, and $w_{ijk}$ are independent standard normal random variables.  The random effects were generated independently as above with the additional interaction random effect $F_\gamma=N(0,\dsiggama^2)$ with  $\dsiggama^2=36$. 
 We then computed the response using the model
\begin{align*}
 y_{ijk}=&\xddd\xi_0+(\xidd-\xddd)\xi_1+(\xdjd-\xddd)\xi_2+(\xijd-\xidd-\xdjd+\xddd)\xi_3+(x_{ijk}-\xijd)\xi_4    \\&
 +\alpha_i+\beta_j+\gamma_{ij}+e_{ijk},\qquad i=1,\ldots,g,\, j=1,\ldots,h,\, k=1,\ldots,m,
\end{align*}
with true parameters $\dot\bxi = [0,5,7,3,4]^T$. We explored 32 scenarios resulting from the combination of the 8 $(g,h,m)$ configurations with the 4 additional random effect settings.

\subsection{Numerical Results}
For each simulation setting in both models, we generated 1000 datasets and fitted the model using REML through the \texttt{lmer} function in the R package \textit{lme4}.  We computed $95\%$ prediction intervals for $\alpha_1$, $\beta_1$ and $\gamma_{11}$ using the normal critical value times the square root of each of the three mean squared error estimates. We examined the empirical coverage (Cvge) and, following \cite{jiang2002unified}, the relative expected length (RLen) of the prediction intervals, where  
\begin{align*}
    \operatorname{RLen}=\frac{\overline{\operatorname{RMSE}}-{\operatorname{RMSE}}_{T}}{{\operatorname{RMSE}}_{T}},
\end{align*}
with $\overline{\operatorname{RMSE}}$ denoting the average of the square roots of the estimates of the mean squared error estimates and  
$\operatorname{RMSE}_T$ the square root of the average over the 1000 simulations of the squared  difference between the EBLUPs (e.g. $\hat\alpha_1$) and the realised value of the random effect (e.g. $\alpha_1$).  
The full set of results are available in the Supplementary material.

\begin{table}[!ht]
\centering
\caption{Simulated coverage and relative length of prediction intervals for random effects, where $\alpha_i$, $\beta_j$ and $e_{ij}$ follow normal distributions.}\label{Tab3}
\begin{tabular}{ccccccccccc}
\toprule
\multicolumn{2}{c}{\multirow{2}{*}{}}                     &      & h=10  &       &  & h=50  &       &  & h=100 &       \\\cline{3-5}\cline{7-8}\cline{10-11}
\multicolumn{2}{c}{}                                      & RMSE & Cvge  & Rlen  &  & Cvge  & Rlen  &  & Cvge  & Rlen  \\\hline
\multirow{9}{*}{$\hat\alpha_1$} & \multirow{3}{*}{g=10}  & LSW   & 0.977 & 0.634 &  & 0.957 & 0.346 &  & 0.939 & 0.27  \\
                                 &                        & KH   & 0.913 & 0.215 &  & 0.929 & 0.23  &  & 0.92  & 0.211 \\
                                 &                        & PR   & 0.947 & 0.384 &  & 0.941 & 0.276 &  & 0.925 & 0.235 \\\cline{2-11}
                                 & \multirow{3}{*}{g=50}  & LSW   & 0.992 & 0.697 &  & 0.964 & 0.36  &  & 0.965 & 0.329 \\
                                 &                        & KH   & 0.939 & 0.22  &  & 0.946 & 0.252 &  & 0.954 & 0.272 \\
                                 &                        & PR   & 0.945 & 0.248 &  & 0.947 & 0.258 &  & 0.956 & 0.276 \\\cline{2-11}
                                 & \multirow{3}{*}{g=100} & LSW   & 0.997 & 0.698 &  & 0.956 & 0.319 &  & 0.968 & 0.336 \\
                                 &                        & KH   & 0.951 & 0.233 &  & 0.943 & 0.214 &  & 0.959 & 0.28  \\
                                 &                        & PR   & 0.952 & 0.246 &  & 0.943 & 0.216 &  & 0.959 & 0.281 \\\hline
                                 &                        &      &       &       &  &       &       &  &       &       \\\hline
\multirow{9}{*}{$\hat\beta_1$}  & \multirow{3}{*}{g=10}  & LSW   & 0.95  & 0.323 &  & 0.96  & 0.317 &  & 0.964 & 0.393 \\
                                 &                        & KH   & 0.914 & 0.212 &  & 0.944 & 0.218 &  & 0.949 & 0.291 \\
                                 &                        & PR   & 0.926 & 0.239 &  & 0.945 & 0.226 &  & 0.949 & 0.295 \\\cline{2-11}
                                 & \multirow{3}{*}{g=50}  & LSW   & 0.931 & 0.272 &  & 0.952 & 0.262 &  & 0.959 & 0.302 \\
                                 &                        & KH   & 0.925 & 0.249 &  & 0.95  & 0.242 &  & 0.952 & 0.281 \\
                                 &                        & PR   & 0.926 & 0.252 &  & 0.95  & 0.243 &  & 0.952 & 0.283 \\\cline{2-11}
                                 & \multirow{3}{*}{g=100} & LSW   & 0.926 & 0.254 &  & 0.951 & 0.283 &  & 0.955 & 0.262 \\
                                 &                        & KH   & 0.924 & 0.243 &  & 0.951 & 0.272 &  & 0.953 & 0.252 \\
                                 &                        & PR   & 0.924 & 0.244 &  & 0.951 & 0.272 &  & 0.953 & 0.252
\\ \bottomrule
\end{tabular}
\end{table}

\begin{table}[!ht]
\centering
\caption{Simulated coverage and relative length of prediction intervals for random effects, where $\alpha_i$ and $e_{ij}$ follow normal distributions, while $\beta_j$ follows a mixture distribution.}\label{Tab4}
\begin{tabular}{ccccccccccc}
\toprule
\multicolumn{2}{c}{\multirow{2}{*}{}}                     &      & h=10  &       &  & h=50  &       &  & h=100 &       \\\cline{3-5}\cline{7-8}\cline{10-11}
\multicolumn{2}{c}{}                                      & RMSE & Cvge  & Rlen  &  & Cvge  & Rlen  &  & Cvge  & Rlen  \\\hline
\multirow{9}{*}{$\hat\alpha_1$} & \multirow{3}{*}{g=10}  & LSW   & 0.986 & 0.657 &  & 0.958 & 0.304 &  & 0.952 & 0.316 \\
                                 &                        & KH   & 0.92  & 0.222 &  & 0.935 & 0.189 &  & 0.939 & 0.254 \\
                                 &                        & PR   & 0.957 & 0.403 &  & 0.946 & 0.229 &  & 0.943 & 0.272 \\\cline{2-11}
                                 & \multirow{3}{*}{g=50}  & LSW   & 0.989 & 0.704 &  & 0.98  & 0.398 &  & 0.954 & 0.239 \\
                                 &                        & KH   & 0.926 & 0.231 &  & 0.961 & 0.286 &  & 0.935 & 0.186 \\
                                 &                        & PR   & 0.939 & 0.26  &  & 0.963 & 0.293 &  & 0.936 & 0.19  \\\cline{2-11}
                                 & \multirow{3}{*}{g=100} & LSW   & 0.993 & 0.762 &  & 0.963 & 0.299 &  & 0.957 & 0.297 \\
                                 &                        & KH   & 0.959 & 0.275 &  & 0.941 & 0.196 &  & 0.948 & 0.242 \\
                                 &                        & PR   & 0.96  & 0.289 &  & 0.942 & 0.198 &  & 0.948 & 0.244 \\\hline
                                 &                        &      &       &       &  &       &       &  &       &       \\\hline
\multirow{9}{*}{$\hat\beta_1$}  & \multirow{3}{*}{g=10}  & LSW   & 0.958 & 0.523 &  & 0.979 & 0.577 &  & 0.971  & 0.486 \\
                                 &                        & KH   & 0.914 & 0.22  &  & 0.946 & 0.321 &  & 0.944 & 0.257 \\
                                 &                        & PR   & 0.934 & 0.337 &  & 0.95  & 0.337 &  & 0.944 & 0.264 \\\cline{2-11}
                                 & \multirow{3}{*}{g=50}  & LSW   & 0.951 & 0.319 &  & 0.946 & 0.252 &  & 0.953 & 0.283 \\
                                 &                        & KH   & 0.929 & 0.228 &  & 0.937 & 0.202 &  & 0.943 & 0.234 \\
                                 &                        & PR   & 0.939 & 0.257 &  & 0.938 & 0.205 &  & 0.943 & 0.236 \\\cline{2-11}
                                 & \multirow{3}{*}{g=100} & LSW   & 0.952 & 0.258 &  & 0.945 & 0.265 &  & 0.958 & 0.213 \\
                                 &                        & KH   & 0.937 & 0.215 &  & 0.943 & 0.238 &  & 0.947  & 0.189 \\
                                 &                        & PR   & 0.94  & 0.225 &  & 0.943 & 0.24  &  & 0.947  & 0.19
\\ \bottomrule
\end{tabular}
\end{table}

Tables~\ref{Tab3} and \ref{Tab4} show results 
for 9 combinations of $(g, h)$ across one normal setting and one mixture setting under model (\ref{two-way cross model no interaction}).
The standard errors for the coverage probabilities obtained as $\{\text{Cvge}(1-\text{Cvge})/1000\}^{1/2}$ are approximately 0.005. These simulation results show that our prediction interval has excellent performance even when $g=h=10$, a very small case for asymptotics based on $g$ and $h$ going to infinity, and illustrate the applicability of our results. 
The coverage results for all three methods are generally good; the relative expected lengths of the intervals are all positive, meaning that the estimates of the root mean squared error are generally larger than the empirical root mean squared errors.
We observe the ordering $\text{KH}\le\text{PR}\le\text{LSW}$ in both coverage and length. For small $g$ and $m$, the KH and LSW intervals for $\hat\alpha_1$ and $\hat\beta_1$ show undercoverage; as both $g$ and $h$ increase (along the diagonal of the tables), the coverage of all three three intervals tends to the nominal level, confirming the large $g$ and $h$ asymptotic results. 
For $\hat\alpha_1$, increasing $g$ with $h$ fixed improves the coverage of the KH and PR intervals, with the coverage approaching the nominal level. Meanwhile, the LSW interval for $\hat\alpha_1$ remains slightly conservative when $h$ is small, but the coverage nears the nominal level as $h$ increases. In the case of $\hat\beta_1$, increasing $h$ with $g$ fixed improves the coverage of the KH and PR intervals,  with the coverage approaching the nominal level. Similarly, the LSW interval for $\hat\beta_1$ remains slightly conservative with small $g$ values but aligns closer to the nominal level as $g$ increases.
The overall conclusion is that our proposed asymptotic prediction interval  consistently shows strong coverage performance, maintaining a close match to the nominal rate and remaining robust even in a small sample size such as $g$ and $h$ both equal to 10.

\begin{table}[!ht]
\centering
\caption{Simulated coverage and relative length of prediction intervals for random effects, where $\alpha_i$, $\beta_j$, $\gamma_{ij}$ and $e_{ijk}$ follow normal distributions.}\label{Tab1}
\begin{tabular}{ccccccccccc}
\toprule
\multicolumn{2}{c}{\multirow{3}{*}{RMSE}} & \multicolumn{4}{c}{m=10}                            &  & \multicolumn{4}{c}{m=30}                            \\\cline{3-6}\cline{8-11}
\multicolumn{2}{c}{}                    & \multicolumn{2}{c}{h=10} & \multicolumn{2}{c}{h=50} &  & \multicolumn{2}{c}{h=10} & \multicolumn{2}{c}{h=50} \\ \cline{3-6}\cline{8-11}
\multicolumn{2}{c}{}                    & Cvge        & Rlen       & Cvge        & Rlen       &  & Cvge        & Rlen       & Cvge        & Rlen       \\\hline
\multirow{2}{*}{$\hat\alpha_1$}    & g=10 & 0.953       & 0.322      & 0.95        & 0.289      &  & 0.962       & 0.403      & 0.933       & 0.269      \\
                               & g=50 & 0.965       & 0.346      & 0.945       & 0.25       &  & 0.972       & 0.414      & 0.947       & 0.218      \\\hline
\multirow{2}{*}{$\hat\beta_1$}     & g=10 & 0.917       & 0.222      & 0.949       & 0.218      &  & 0.946       & 0.347      & 0.96        & 0.301      \\
                               & g=50 & 0.929       & 0.583      & 0.942       & 0.251      &  & 0.925       & 0.26       & 0.952       & 0.293      \\\hline
\multirow{2}{*}{$\hat\gamma_{11}$} & g=10 & 0.967       & 0.395      & 0.98        & 0.425      &  & 0.972       & 0.368      & 0.972       & 0.343      \\
                               & g=50 & 0.967       & 0.438      & 0.969       & 0.349      &  & 0.963       & 0.395      & 0.959       & 0.342     \\
\bottomrule
\end{tabular}
\end{table}
\begin{table}[!ht]
\centering
\caption{Simulated coverage and relative length of prediction intervals for random effects, where $\alpha_i$ and $\gamma_{ij}$ follow normal distributions, while $\beta_j$ and $e_{ijk}$ follow mixture distributions.}\label{Tab2}
\begin{tabular}{ccccccccccc}
\toprule
\multicolumn{2}{c}{\multirow{3}{*}{RMSE}} & \multicolumn{4}{c}{m=10}                            &  & \multicolumn{4}{c}{m=30}                            \\\cline{3-6}\cline{8-11}
\multicolumn{2}{c}{}                    & \multicolumn{2}{c}{h=10} & \multicolumn{2}{c}{h=50} &  & \multicolumn{2}{c}{h=10} & \multicolumn{2}{c}{h=50} \\ \cline{3-6}\cline{8-11}
\multicolumn{2}{c}{}                    & Cvge        & Rlen       & Cvge        & Rlen       &  & Cvge        & Rlen       & Cvge        & Rlen       \\\hline
\multirow{2}{*}{$\hat\alpha_1$}     & g=10  & 0.958       & 0.402      & 0.939       & 0.298      &  & 0.973       & 0.375      & 0.941       & 0.313      \\
                                & g=50  & 0.973       & 0.449      & 0.957       & 0.253      &  & 0.975       & 0.473      & 0.947       & 0.306      \\\hline
\multirow{2}{*}{$\hat\beta_1$}      & g=10  & 0.947       & 0.348      & 0.955       & 0.300      &  & 0.939       & 0.347      & 0.968       & 0.356      \\
                                & g=50  & 0.947       & 0.308      & 0.951       & 0.262      &  & 0.937       & 0.233      & 0.961       & 0.237      \\\hline
\multirow{2}{*}{$\hat\gamma_{11}$}  & g=10  & 0.972       & 0.447      & 0.963       & 0.335      &  & 0.962       & 0.382      & 0.961       & 0.349      \\
                                & g=50  & 0.972       & 0.485      & 0.960       & 0.365      &  & 0.970       & 0.450      & 0.947       & 0.267     \\
\bottomrule
\end{tabular}
\end{table}

Tables~\ref{Tab1} and \ref{Tab2} show results for 8 combinations of $(g, h, m)$ across one normal setting and one mixture setting under model (\ref{two-way cross model}). 
The standard errors for the coverage probabilities obtained as $\{\text{Cvge}(1-\text{Cvge})/1000\}^{1/2}$ are approximately 0.006. These simulation results show that our prediction interval has excellent performance even when $g=h=m=10$, a small case for asymptotics based on $g, h$ and $m$ going to infinity, and illustrate the applicability of our results. The relative expected lengths of the intervals are all positive, meaning that the estimates of the root mean squared error are generally larger than the empirical root mean squared errors.

\section{Real data Analysis}\label{sec7: real data}
We illustrate the use of our prediction intervals for random effects in model (\ref{two-way cross model no interaction}) using movie data described in \cite{harper2015movielens}. The movie dataset is obtained from Kaggle  at \url{https://www.kaggle.com/datasets/rounakbanik/the-movies-dataset}. In a full analysis of these data, we would fit and compare different regression models to understand the data. However, as our purpose is to illustrate our theoretical results, we just fit one model to a subset of the data. We selected a subset of movie ratings from $g = 126$ customers for $h = 66$ movies, yielding $n = 8316$ ratings. The response $y_{ij}$ represents the satisfaction rating of customer $i$ for movie $j$, made on a five-point scale with half-star increments (0.5 stars to 5.0 stars). The data include features of customers and movies. We model the relationship between the ratings and four explanatory variables: movie popularity ({\em Popularity}), movie revenue ({\em Revenue}), movie runtime ({\em Runtime}), and the time gap in days between the movie release date and the customer rating date ({\em DayGap}). Writing the cell-level covariate {\em DayGap} as $x_{ij}$, we decompose $x_{ij}$ into $x_{ij}-\bar{x}_{i.} - \bar{x}_{.j} + \bar{x}$, $\bar{x}_{i.} - \bar{x}$, and $\bar{x}_{.j} - \bar{x}$ (plus $\bar{x}$ which we absorb into the intercept).   We label these components by adding {\em cell\_cent}, {\em row\_cent}, and {\em column\_cent} to the variable name, respectively. We model the movie rating results by fitting a two-way crossed random effect model with no interaction, given by
\begin{align*}
y_{ij} = & \xi_0 + \xi_1 \text{DayGap\_row\_cent}_i + \xi_2 \text{DayGap\_column\_cent}_j + \xi_3 \text{Popularity}_j \\
& + \xi_4 \text{Revenue}_j + \xi_5 \text{Runtime}_j + \xi_6 \text{DayGap\_cell\_cent}_{ij} + \alpha_i + \beta_j + e_{ij},
\end{align*}
where the independent random variables $\alpha_i$, $\beta_j$, and $e_{ij}$ represent the customer random effects, movie random effects, and error terms, respectively. The movie random effects represent an intrinsic rating of the movies (after adjustment for the fixed effects) so are of primary interest; the contributions of the fixed effects are also of interest.  We fitted the model using the \texttt{lmer} function from the \texttt{lme4} package.

\begin{table}[!h]

\caption{Parameter estimates (REML) for the Movie Rating Data}\label{Tab 5}
\centering
\begin{tabular}{lccl}
\toprule
Effect                   & Estimate & Std Error& $p$-value \\\hline
{\em (Intercept)}            & \phantom{1}3.818  & 0.059  & $9.03 \times 10^{-98}$ (${\ast\ast}$)   \\
{\em DayGap\_row\_cent}       & -0.153   & 0.036     & $3.33 \times 10^{-5}$ (${\ast\ast}$)\\
{\em DayGap\_column\_cent}    & \phantom{1}0.0127 & 0.049 & $1.26 \times 10^{-2}$ (${\ast}$)\\
{\em Popularity}          & \phantom{1}0.129    & 0.050   & $1.25 \times 10^{-2}$ (${\ast}$)   \\
{\em Revenue}                & -0.048   & 0.051   & $3.58 \times 10^{-1}$  \\
{\em Runtime}                & -0.060    & 0.052  & $2.52 \times 10^{-1}$   \\
{\em DayGap\_cell\_cent}     & -0.037   & 0.008  & $5.13 \times 10^{-6}$ (${\ast\ast}$)    \\\hline
Customer $\sigma_\alpha$ & \phantom{1}0.388    &          & \\
Movie $\sigma_\beta$   & \phantom{1}0.379    &           &\\
Rating $\sigma_e$        & \phantom{1}0.728    &          &\\ \bottomrule
\multicolumn{4}{l}{
Significance levels: $\ast\ast$ ($p < 0.01$), $\ast$ ($p < 0.05$), no symbol ($p \geq 0.1$).
}\\
\multicolumn{4}{l}{NOTE:  \texttt{lmer} does not compute Std Error for the variance components.}
\end{tabular}
\end{table}
\begin{figure}[!h]
	\includegraphics[width=0.5\linewidth]{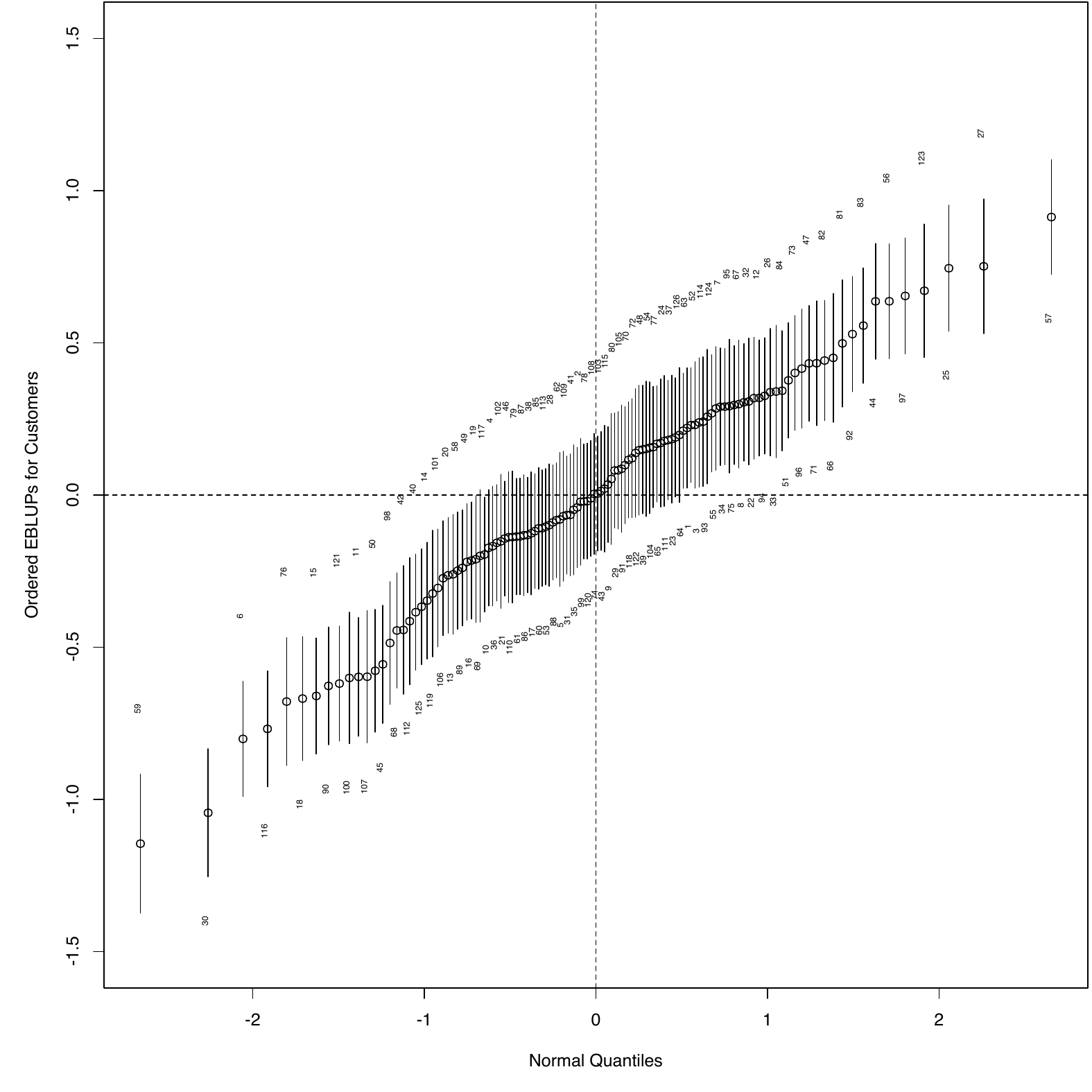}
 	\includegraphics[width=0.5\linewidth]{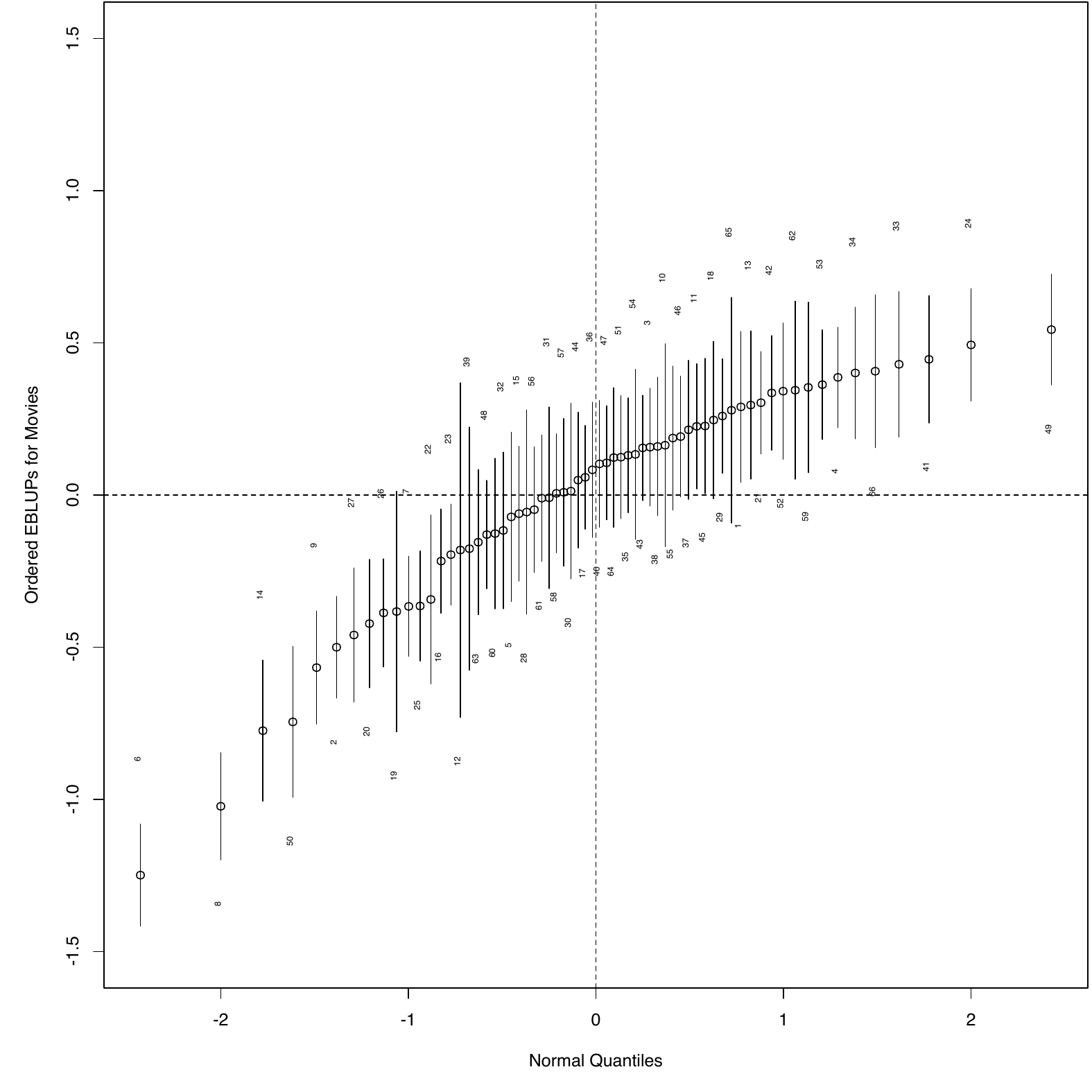}
	\caption{Asymptotic prediction intervals for the individual random effects representing customer effects and movie effects in two-way crossed random effect model for movie ratings with no interaction. The prediction intervals are superimposed on quantile-quantile plots of the point predictions.}\label{fig 1}
\end{figure}
The  REML estimates of the parameters are shown in Table~\ref{Tab 5}. With the exception of {\em Revenue} and {\em Runtime}, the covariates in the model are significant. Interestingly, the {\em DayGap} terms at the different levels have different effects.
We then computed the EBLUPs for the random effects, our estimates of their prediction mean squared errors, and $95\%$ prediction intervals for the  customer $\alpha_i$ and movie $\beta_j$ random effects, respectively. 
We present these intervals for the random effects on a normal qq-plot of the EBLUPS with each EBLUP labeled by the corresponding customer and movie.  The customer random effects are plausibly normally distributed, but the movie random effects have an asymmetric distribution with a short upper tail: the lowest rated movies are clearly identified, but the highest-rated movies are less distinguishable.  The top three movies by estimated random effect are numbers 49, 24 and 41.
 
\section{Conclusion}\label{sec8: discussion}

\cite{lyu2024increasing} recently established a simple central limit theorem for the maximum likelihood and REML estimators of the parameters in crossed random effect models under simple moment conditions which allow nonnormality in the random effects and/or errors, and allow the number of rows and columns (and cell sizes when fitting interactions) to tend to infinity without any constraints on their relative rates of convergence. 
In this paper, we use these results to describe the joint asymptotic behavior of EBLUPs of the unobserved individual and cell-level random effects in crossed random effect models, both with and without interaction. Specifically, we obtain the asymptotic distribution of the EBLUPs and show that this distribution is approximately the convolution of the true distribution of the random effect and a normal distribution that depends on the number of rows, columns, and cell sizes. This new result provides valuable insight into the asymptotic behavior of EBLUPs.

Our theoretical results lead to straightforward approximations for the MSE of EBLUPs, which are easy to estimate and use for constructing asymptotic prediction intervals for the random effects. Although prediction intervals have been used for random effects, this is the first time these intervals have been properly justified in crossed random effect models. Our approximation and estimator of the MSE of the EBLUPs are much simpler than those of \cite{kackar1984approximations} and \cite{prasad1990estimation} and the asymptotic properties of their estimators under crossed-models are not known.

In contrast, we work within an asymptotic framework where EBLUPs are consistent. We make a linear approximation to the EBLUPs and then compute the MSE. The advantage of our approach is its simplicity (in conditions and in execution).  Additionally, in our approach, BLUPs and EBLUPs have the same asymptotic distribution, aligning them with methods like generalized least squares for estimating regression parameters, where the estimators share the same asymptotic distribution whether the variance parameters are known or estimated consistently. We also present a simulation study to demonstrate the good performance of both our approximation and its estimator in finite samples.

\bibliographystyle{agsm}
\bibliography{mythesisbib}
\end{document}

%% file: definition.tex
\newtheorem{theoremm}{Theorem} 

\newtheorem{corollary}[theoremm]{Corollary}

\newcommand{\E}{\operatorname{E}}

\newcommand{\var}{\operatorname{Var}}

\newcommand{\tr}{\operatorname{tr}}
\newcommand{\diag}{\operatorname{diag}}

\newcommand{\bB}{\mathbf{B}}
\newcommand{\bC}{\mathbf{C}}
\newcommand{\bD}{\mathbf{D}}

\newcommand{\bG}{\mathbf{G}}

\newcommand{\bI}{\mathbf{I}}
\newcommand{\bJ}{\mathbf{J}}
\newcommand{\bK}{\mathbf{K}}

\newcommand{\bP}{\mathbf{P}}

\newcommand{\bV}{\mathbf{V}}

\newcommand{\bX}{\mathbf{X}}

\newcommand{\bZ}{\mathbf{Z}}

\newcommand{\bd}{\mathbf{d}}
\newcommand{\be}{\mathbf{e}}

\newcommand{\bu}{\mathbf{u}}

\newcommand{\bx}{\mathbf{x}}
\newcommand{\by}{\mathbf{y}} 

\newcommand{\balpha}{\boldsymbol{\alpha}}
\newcommand{\btheta}{\boldsymbol{\theta}}
\newcommand{\bomega}{\boldsymbol{\omega}}

\newcommand{\bgamma}{\boldsymbol{\gamma}}

\newcommand{\bphi}{\boldsymbol{\phi}}
\newcommand{\bbeta}{\boldsymbol{\beta}}

\newcommand{\bone}{\boldsymbol{1}}
\newcommand{\bxi}{\boldsymbol{\xi}}

\newcommand{\bOmega}{\boldsymbol{\Omega}}

\newcommand{\siga}{{\sigma}_{\alpha}}  
\newcommand{\sigb}{{\sigma}_{\beta}}  
\newcommand{\siggama}{{\sigma}_{\gamma}}  
\newcommand{\sige}{{\sigma}_{e}}  

\newcommand{\dsiga}{\dot{\sigma}_{\alpha}}  
\newcommand{\dsigb}{\dot{\sigma}_{\beta}}  
\newcommand{\dsiggama}{\dot{\sigma}_{\gamma}}  
\newcommand{\dsige}{\dot{\sigma}_{e}}